# Towards Robust and Scalable Dispatch Modeling of Long-Duration Energy Storage

Omar J. Guerra [a], Sourabh Dalvi [a], Amogh Thatte [b], Brady Cowiestoll [a], Jennie Jorgenson [a], and Bri-Mathias Hodge [a, c, d]

[a] National Renewable Energy Laboratory, 15013 Denver West Parkway, Golden, CO 80401, USA
[b] Colorado School of Mines – Advanced Energy Systems Graduate Program, 1500 Illinois Street, Golden, CO 80401, USA.
[c] Renewable and Sustainable Energy Institute, University of Colorado, Boulder, CO 80309, USA
[d] Department of Electrical, Computer & Energy Engineering, University of Colorado, Boulder, CO 80309, USA

## Abstract

Energy storage technologies, including short-duration, long-duration, and seasonal storage, are seen as technologies that can facilitate the integration of larger shares of variable renewable energy, such as wind and solar photovoltaics, in power systems. However, despite recent advances in the techno-economic modeling of energy storage (particularly for short-duration applications), the operation and economics of long-duration energy storage are still incomplete in power systems modeling platforms. For instance, existing modeling approaches for long-duration storage are often based either on an oversimplified representation of power system operations or limited representation of storage technologies, e.g., evaluation of only a single application. This manuscript presents an overview of the challenges of modeling long-duration energy storage technologies, as well as a discussion regarding the capabilities and limitations of existing approaches. We used two test power systems with high shares of both solar photovoltaics- and wind (70% - 90% annual variable renewable energy shares) to assess long-duration energy storage dispatch approaches. Our results estimate that better dispatch modeling of long-duration energy storage could increase the associated operational value by 4% - 14% and increase the standard capacity credit by 14% - 34%. Thus, a better long-duration energy storage dispatch could represent significant cost saving opportunities for electric utilities and system operators. In addition, existing long-duration dispatch modeling approaches were tested in terms of both improved system value (e.g., based on production cost and standard capacity credit) and scalability (e.g., based on central processing unit time and peak memory usage). Both copper plate and nodal representations of the power system were considered. Although the end volume target dispatch approach, i.e., based on mid-term scheduling, showed promising performance in terms of both improved system value and scalability, there is a need for robust and scalable dispatch approaches for long-duration energy storage in transmission-constrained electric grids. Moreover, more research is required to better understand the optimal operation of long-duration storage considering extreme climate/weather events, reliability applications, and power system operational uncertainties.

## Highlights

- Long-duration energy storage dispatch approaches are reviewed.
- Performance of energy storage dispatch approaches is assessed.
- A novel metric for energy storage capacity credit estimation is proposed.
- Future research directions for modeling the dispatch of energy storage are discussed.

## Nomenclature

**List of Indices**

| Index | Definition |
|---|---|
| $t$ | Index of time periods, where $t \in \{1,2,\ldots,NT\}$ |
| $i$ | Index of generation units, where $i \in \{1,2,\ldots,NU\}$ |
| $s$ | Index of storage units, where $s \in \{1,2,\ldots,NS\}$ |
| $j, k$ | Index of nodes where $j, k \in \{1,2,\ldots,J\}$ |
| $r$ | Index of reserve type where $r \in \{Reg\ Up, Flex/Spin\ Up\}$ |

**List of parameters**

| Parameter | Definition |
|---|---|
| $C^{fuel}, C^{start}, C^{stop}$ | Cost of fuel, start up, and shutdown |
| $P^{min}, P^{max}$ | Minimum and maximum dispatch of generation unit |
| $R^{up}, R^{down}$ | Ramp-up and down limit for generation unit |
| $RR_{r,t}$ | Reserve requirement '$r$' at time '$t$' |
| $F^{min}, F^{max}$ | Minimum and maximum line power-flow limit |
| $B$ | Susceptance of transmission line |
| $PC^{max}, PD^{max}$ | Maximum charging and discharging limit of the storage |
| $\theta^{min}, \theta^{max}$ | Minimum and maximum power flow angle |
| $SoC^{min}, SoC^{max}$ | Minimum and maximum state-of-charge for the storage |
| $\eta^{sd}, \eta^{c}, \eta^{d}$ | Self-discharge, charging, and discharging efficiency of the storage |

**List of variables**

| Variable | Definition |
|---|---|
| $p_{i,t}$ | Power produced by generator '$i$' at time '$t$' |
| $p_{s,t}^{d}$ | Power discharged by storage '$s$' at time '$t$' |
| $p_{s,t}^{c}$ | Power charged by storage '$s$' at time '$t$' |
| $r_{r,\ i/s,\ t}$ | Reserve '$r$' provided by generator '$i$' or storage '$s$' at time '$t$' |
| $d_{k,t}$ | Demand/load at node '$k$' at time '$t$' |
| $f_{t,j,k}$ | Power flowing between nodes '$j$' and '$k$' at time '$t$' |
| $SoC_{s,t}$ | State-of-charge of storage '$s$' at time '$t$' |
| $x_{i,t}$ | Binary variable for tracking commitment status of generation unit '$i$' at time '$t$' |
| $x_{i,t}^{start}, x_{i,t}^{stop}$ | Binary variables for tracking startup and shutdown status of generation unit '$i$' at time '$t$' |
| $x_{s,t}^{c}$ | Binary variable representing charging of storage '$s$' at time '$t$'. Discharging is represented by $(1 - x_{s,t}^{c})$ |

# 1 Introduction

A growing interest in reducing emissions from the electricity sector, as well as cost reductions in variable renewable energy (VRE) generation technologies such as solar photovoltaic (PV) and wind power, have resulted in increased shares of renewable energy generation in the United States and across the globe [1,2]. Cost declines for many types of energy storage technologies have also driven an interest in assessing the role of storage as a grid asset and in assisting the integration of renewables [3]. Lithium-ion battery technologies have shown the most dramatic cost declines, indicating that near-term deployment of grid-scale storage technologies will likely be dominated by batteries of relatively short duration, e.g., less than 10 hours[4–6]. However, as renewable energy and short-duration battery technologies continue to enter the market, and as deeper decarbonization continues to be a priority, longer-duration storage technologies may look increasingly attractive to address longer-term mismatches in supply and demand [3,7–12], as illustrated in Figure 1. Longer-duration storage may also provide higher capacity value —associated with the ability of the storage device to supply energy during periods of peak (net) load— or contribution to a resource adequacy requirement [13]. As power systems move towards more decarbonization futures, it will become increasingly important for utilities to understand the value of long-duration energy storage (LDES) for their changing systems[8,14]. This manuscript describes some of the challenges of modeling the dispatch of LDES, as well as some of the approaches that have been proposed to address this decision-making problem. The performance of different proposed LDES dispatch approaches is also assessed in terms of different economic, operation, and computational metrics for two test power systems with high shares of VRE.

It is difficult to define a precise energy capacity range for what constitutes LDES. A review of approximately forty papers finds that most literature seems to be converging on a definition of an energy to power ratio of 10-100 hours [15,16]. This typically captures a range that may be operated on a timescale longer than diurnally, but not fully seasonal in nature. The issue of inter-temporal storage dispatch scheduling is similarly complicated for both long-duration and seasonal storage [17–21]. The suite of technology options for new storage with longer durations are less clear but could consist of longer-duration versions of existing battery technologies, new pumped-storage development, thermal energy storage, or some versions of power-to-gas technologies [4,5,22,23].

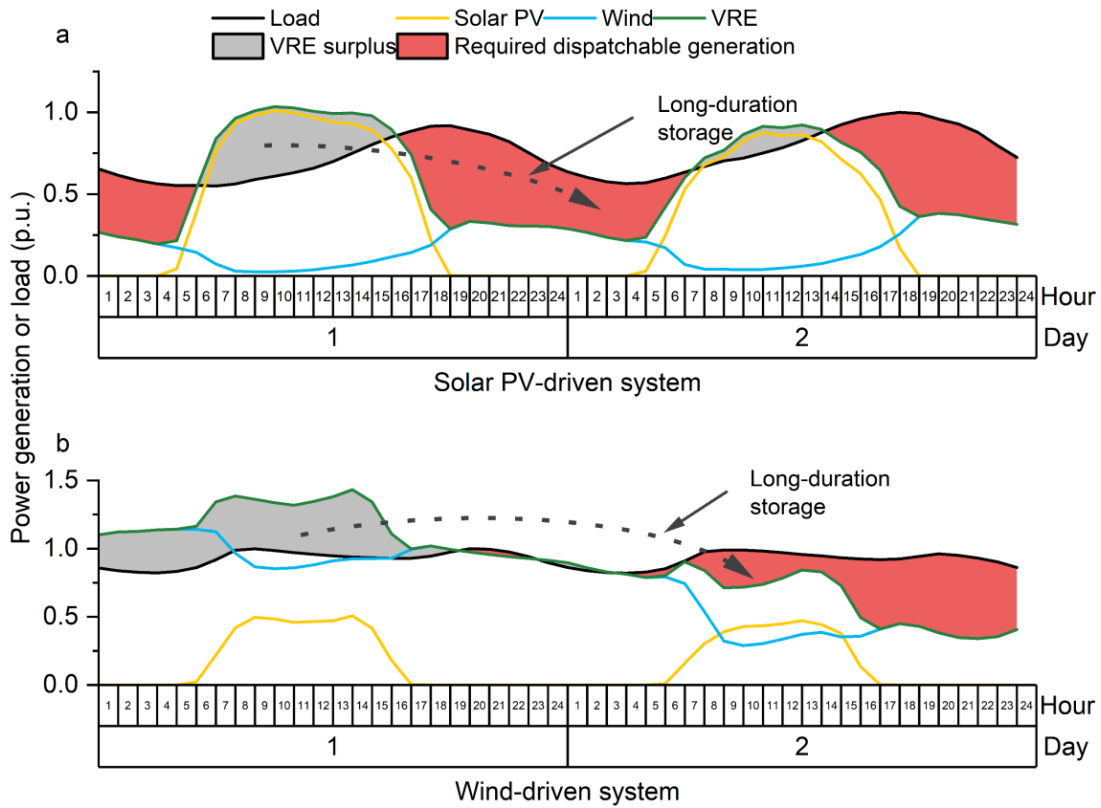

Figure 1. The role of long-duration for the integration of high shares of variable renewable energy (VRE) sources in solar PV-driven (a) and wind-driven (b) power systems. Data for the solar PV-driven and wind-driven power systems are based on CAISO [California Independent System Operator] and MISO [Midcontinent Independent System Operator] data, respectively[24]. The data were normalized based on the peak load for the two-day time frame.

The operational synergies between solar PV and diurnal storage, with < 6 hours duration [15], are clear given the predictable daily on-off cycle of solar PV; storage charges during the day when the sun is shining and generates during the evening or morning load ramps when solar PV is not available [25]. However, questions remain regarding optimal dispatch strategies for LDES. One key consideration is the decision-making process around dispatching stored energy for intra-day energy arbitrage versus reserving it for times of potential higher value further in the future (e.g., inter-day energy arbitrage or potential reliability needs) and forecasting whether more advantageous times to charge or dispatch storage will occur within a given time frame. Production cost models (PCMs) are tools that can evaluate such trade-offs by considering the optimal unit commitment and economic dispatch of the generators in an electricity system, including storage. Due to computational complexity and to align with day-ahead energy markets, PCMs are often formulated to consider one to two days at a time and are run daily, which limits the central processing unit (CPU) time allowed for solving PCMs. As a result, there is a need to consider the potential value of storage outside of the PCM horizon to fully capture the opportunity values of inter-temporal storage operations, an area of growing research [26,27] while maintaining computational tractability. For example, if a PCM or similar model considers only a day (or two) in an optimization problem, there is no value in saving the energy beyond that horizon without additional information[28]. Simpler models such as price-taker models (which assume that a storage device will not materially impact the local energy prices) can often consider longer time horizons due to the relatively simplistic optimization formulation[21,29]. The results of price-taker models can then feed more complex models such as PCMs. One such analysis of high renewable systems in the western United States found additional value with long-duration storage compared to short-duration (in this case, 8 hours), but that the vast majority (80%-90%) of storage value is realized by short-duration storage [29], this is also true for other systems[30–32].

Other studies have attempted to develop dispatch algorithms for storage that optimize the financial benefits to the owner of the device, e.g., based on price taker-models [33,34]; such approaches focus on single devices and not the entire power sector and so may not be computationally scalable to full system optimization. Similarly, an analysis of coupled operation of a solar PV- pumped hydro storage (PHS) plant has been performed [35], optimizing for self-sufficiency of the coupled system but not for system-wide operations. Optimizing storage dispatch to maximize financial benefits to the device owner may not be the most beneficial from a system point of view, considering possible market inefficiencies and the inability of market structures to consider all possible value streams for storage [36].

LDES has also been studied across many regions and scales, though primarily in planning-level capacity expansion models [37,38]. For instance, a previous study showed that up to 80% carbon-free or renewable energy mix in U.S. power systems could be achieved with economic curtailment and a portfolio of short- and LDES [24]. Then, depending on the power system, there is a point between 80% and 95% when seasonal storage becomes cost-effective, if a higher decarbonization or renewable integration goal is to be met[24]. Previous studies have shown that the required deployment of storage power capacity increases linearly while the required storage energy capacity, i.e., duration, increases exponentially with the deployment of wind and solar PV power sources [24,39]. Optimal siting and sizing of storage deployment is critical to fully realize the power system benefits of energy storage technologies. For instance, if properly located, storage technologies have the potential to reduce the need for transmission capacity expansion [40]. To this end, storage deployment tools should include more detailed spatial resolution and representation of transmission investment options [27]. Recent reviews have summarized the challenges of modeling energy storage in long-term power planning models, including the diversity of energy storage technologies and the need for higher temporal and spatial resolutions, among others [27,41,42]. Capacity expansion models typically optimize the entire year (or multiple years) simultaneously based on representative time slices without preserving chronology. Although capacity expansion models are helpful for investment decision making and understanding the value such technologies might be able to bring to the power system, the current temporal granularity and the inability to accurately track the state-of-charge (SOC) of storage does not provide the level of realism needed to fully assess the potential value of long-duration energy storage. Additionally, increasing the temporal and spatial granularity of capacity expansion models to better represent the operation of energy storage could significantly increase the associated computation burden, so specialized solution techniques could be required (e.g., decomposition and/or parallelization methods). Note that there are capacity expansion models that do use full chronology, if needed, but the associated computational cost imposes limits, such as looking at a smaller geographic region, fewer technologies, or fewer scenarios.

The challenges and the assessment of proposed approaches for modeling the operational dispatch of LDES are the focus of this manuscript. In the next section, an overview of the challenges and proposed modeling approaches for the dispatch of LDES are presented, considering the integration of large shares of VRE. Subsequently, the performance of different proposed dispatch approaches for long-duration storage are tested and quantified in terms of economic, operational, and computational metrics for two test systems with high VRE shares. Finally, conclusions are drawn, and future research directions are discussed.

## 2 Modeling of long-duration energy storage

The modeling of LDES represents a major challenge in power system operations, including for the most commonly employed system optimization models. First, the modeling of energy storage requires high temporal resolution, e.g., at least hourly, to accurately account for short-term variability of wind and solar PV power generation and the corresponding interactions with storage operations, e.g., intra-day shifting of wind and solar PV generation [27]. Additionally, the mismatch between load and wind and solar PV generation also occurs beyond the intra-day timescale, e.g., at inter-day and seasonal timescales [26]. Thus, modeling energy storage over longer optimization windows (weeks, months, or year) is required to fully assess multiscale energy storage services, such as balancing inter-day and seasonal mismatch between load and VRE generation, firm capacity, operating reserves, and reliability services [20,26,43]. Moreover, because storage levels are sequential state variables, preserving chronology is critical for the appropriate modeling of energy storage [20,27,44].

There are a variety of potential LDES technologies that are commercially available or under development (e.g., Li-ion batteries, compressed air energy storage, pumped hydro storage, flow batteries, and hydrogen storage) with each technology having a specific system configuration, cost structure, and technological parameters. For example, some technologies can decouple charging and discharging power capacities (e.g., compressed air energy storage and hydrogen) which represents potentially valuable flexibility from the investment and operational viewpoints [24]. Additionally, some technologies could have limited cycling capabilities, e.g., driven by chemical degradation and mechanical ageing of electrochemical systems, which could reduce the flexibility that these technologies could provide to power systems. Moreover, storage devices can be operated in storage-to-storage charging operations — one storage device used to charge another storage device, which can sometimes reduce total power system costs [24]. Thus, the comprehensive modeling of storage technologies should be flexible enough to include a variety of storage technology characteristics such as cycling constraints, lifetime, decoupling of charging and discharging power capacities, and inter-storage operations (e.g., simultaneous assessment of multiscale storage technologies).

Uncertainties in model parameters and imperfections in the mathematical formulation of the optimization problem can also impact the deployment and operation of any technology, including energy storage technologies [20,27]. For example, oversimplified mathematical formulations of energy storage operations may fail to capture real challenges with operating storage, such as requirements for maintaining an average SOC, variations between charging and discharging efficiencies, dissipation of stored energy, or cycle degradation, which could impact how these devices would be operated. Failure to accurately represent all storage parameters could over or underestimate the value these devices bring to the power system. Therefore, comprehensive assessment of energy storage deployment and operation should quantify the effects of these often-overlooked operational aspects.

To this end, different approaches for representing the value of LDES have been implemented, including sensitivity analyses, multiscale load and renewable generation forecasts, price-taking storage dispatch under a power purchase agreement [45], real-time pricing algorithms [33,46], game theory [34], and stochastic optimization [47]. However, the implementation of each of these approaches includes computational challenges when applied at the transmission-level power systems scale. Note that this section is not intended to provide a comprehensive overview of energy storage modeling gaps. Rather, this section summarizes key modeling issues associated with grid-integrated long-duration storage technologies from a central planner perspective. For more detailed examinations of energy storage modeling issues, the reader is referred to recent reviews on this topic [20,27,48], including both the central planner and merchant/developer perspectives. The following sections describe four approaches that have been proposed in the literature to address the modeling of LDES, including some of the pros and cons of each approach. These approaches address some of the modeling issues described previously in this section, but important gaps remain.

## 2.1 Extended optimization horizon or window of foresight

A primary issue when modeling energy storage in day-ahead power system optimization models is the draining of the storage devices at the end of the simulated time frame or optimization horizon, because no value is placed on storing energy for usage outside of that time frame. Many studies include a period of one day for their optimization window [38,49,50]. A simple solution to the problem of end-of-period storage effects would be to extend the optimization horizon to consider more than one day at time. For long-duration storage, this might be several days, a week, or a month. However, increasing the optimization horizon often increases the number of variables and equations in the PCM formulation which can dramatically increase the computation time. Additionally, some long-duration storage may require even longer operational time horizons, such as months. Extending the optimization horizon quickly becomes increasingly difficult to solve on all but the most trivial systems.

Another similarly straightforward approach to valuing storage beyond the optimization time frame is to add some foresight (also often called a look-ahead window) as shown in Figure 2. This approach passes more information to the optimization solver to present a more complete picture of the near-term future value of stored energy. Often, the look-ahead window may be optimized at a less resolved time resolution or include fewer constraints than during the primary optimization horizon to maintain computational tractability. Many studies include one [49–51] or two days [25,38] for this look-ahead window. The end effects can then be removed before passing the SOC to the next time frame or model run [52,53]. One key limitation or condition of this approach is that the look-ahead time frame should be longer than the expected operational time frame of the storage device [52]. Thus, the required look-ahead time frame depends on the energy capacity or duration of the storage technologies to be modeled and system needs. Therefore, this approach may not be practical for modeling long-duration or seasonal storage technologies due to the increased computational requirements. In summary, the capability of this approach to modelling long-duration and seasonal energy storage requires a better fundamental understanding of the computational versus fidelity trade-offs, including a more comprehensive analysis of the impacts of different look-ahead windows on the value of energy storage across different timescales.

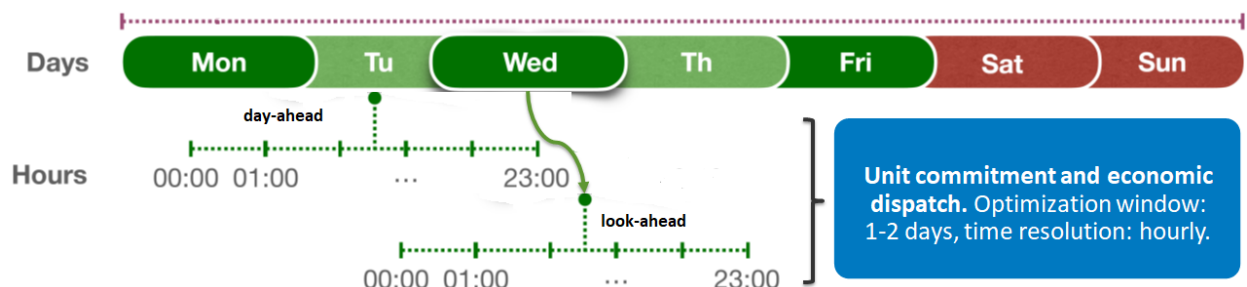

Figure 2. A Look-ahead approach for modeling long-duration energy storage. This example shows a 1-day simulation plus one additional day of foresight, which could include simplifications, e.g., temporal aggregation, relaxation of some power system constraints, etc.

## 2.2 End volume targets

The end volume targets method [21,47] uses an exogenous storage dispatch model (perhaps using a simplified production cost approach or price-taker assumptions) or a stochastic optimization model to optimize the operation of storage technologies in what can be termed a medium- or mid-term (MT) simulation phase. Then, the optimal storage dispatch from the simplified MT phase is input as SOC targets for the standard PCM run. These external targets may apply on a daily, weekly, or even monthly basis depending on the formulation of the PCM run. These targets, which could be based on heuristics, e.g., given percentage of the maximum SOC, are used as constraints to enforce the long-duration or seasonal dispatch behavior of energy storage, while keeping the flexibility to optimize intra-day operation of the storage devices in the unit commitment and economic dispatch model, as illustrated in Figure 3. Note that this approach does not require changes to the standard power system production cost modeling. Optimization of unit commitment and economic dispatch for 1 day-ahead plus 1 day look-ahead, is generally in line with current day-ahead electricity markets. The main drawback to this approach is that the MT production cost, storage dispatch or stochastic models may not accurately represent power system details (transmission constraints, operating reserves, ramp constraints, etc.) which could impact the deployment and operation of energy storage technologies. This approach also requires an additional model solution, potentially increasing total solve times, but has well-documented use cases, particularly in handling hydropower dispatch in the Brazil, Norway, and the western United States[21,54].

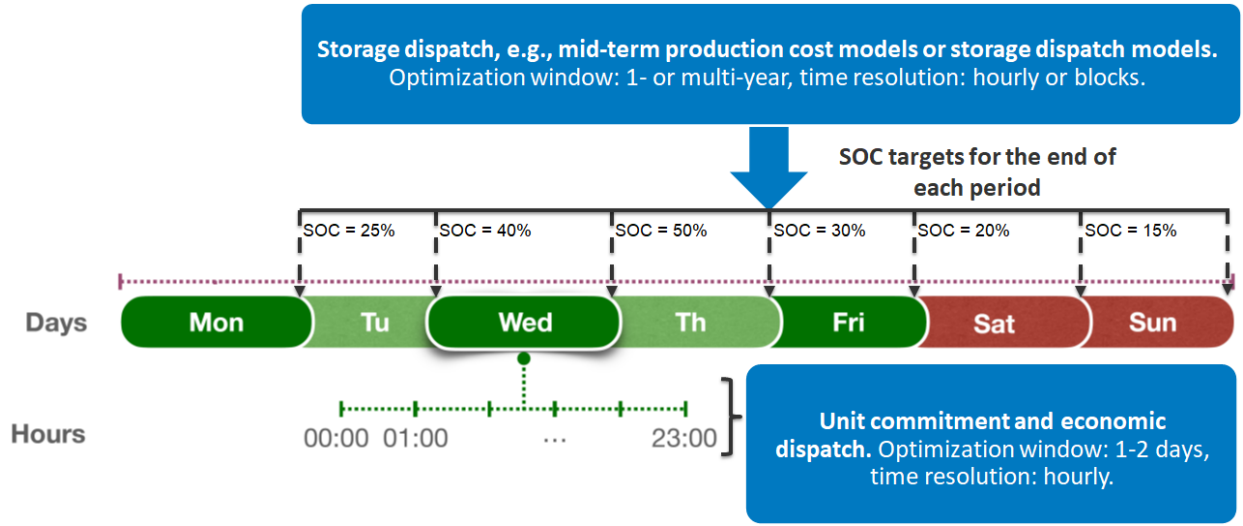

Figure 3. End volume target (e.g., state of charge [SOC] = 25% of max SOC) approach for modeling long-duration and seasonal energy storage (based on Guerra et al. 2020[21]).

A previous study used a multi-model approach for the implementation of the end volume targets method for a system that had long-duration or seasonal storage power capacity designed to represent around 1% of the peak load of the power system, while different energy capacities or discharge durations were evaluated, including 1 day, 2 days, 1 week, 2 weeks, and 1 month[21,29]. A production cost model was used to perform chronological simulation of annual power system operations without any long-duration storage and using hourly temporal resolution and a 1 day-ahead optimization window (365 sequential 1-day, i.e., 24-hour, runs for each year). After this first PCM run, the resulting net load and locational marginal prices (LMPs) for a given region were used to optimize the operation of a long-duration or seasonal storage device using a price-taker model with hourly temporal resolution and a 1-year optimization window. Based on the optimal storage dispatch from the price-taker model, SOC targets were estimated for the end of each day. Finally, the long-duration or seasonal storage device was included in the PCM with end volume targets represented as constraints to enforce the long-duration or seasonal dispatch of the storage device while optimizing the hourly operation of the storage device based on the optimal unit commitment and economic dispatch (second PCM run). The outputs were analyzed to evaluate the corresponding total system value, including operational value and capacity value.

## 2.3 Stored Energy Value

The "stored energy value" (sometimes called "water value"), the marginal future value of storing an additional unit of energy, is another approach for long-duration storage dispatch that does not require any consideration of longer time horizons, nor variable time steps, and is not computationally demanding [55–58]. In simple terms, the methodology applies a value (usually in $/MWh) to the stored energy, as described in Figure 4. If the marginal price of energy is lower than the value of stored energy (accounting for efficiency losses), then the storage device would "charge up" to store the cheaper energy. If the marginal price of energy is higher than the value of stored energy (again accounting for any losses), then the device would discharge. One study considering high amounts of variable generation in the city of Los Angeles utilized this approach [59]. This study found that adding a value to the stored energy encouraged the storage devices to

charge on zero-marginal-cost energy (such as excess generation from wind or solar PV) and encourage a higher SOC at the end of the optimization window even without any foresight. Additionally, a previous study found that the "stored energy value" approach has shorter run times in comparison with the end volume targets approach [58].

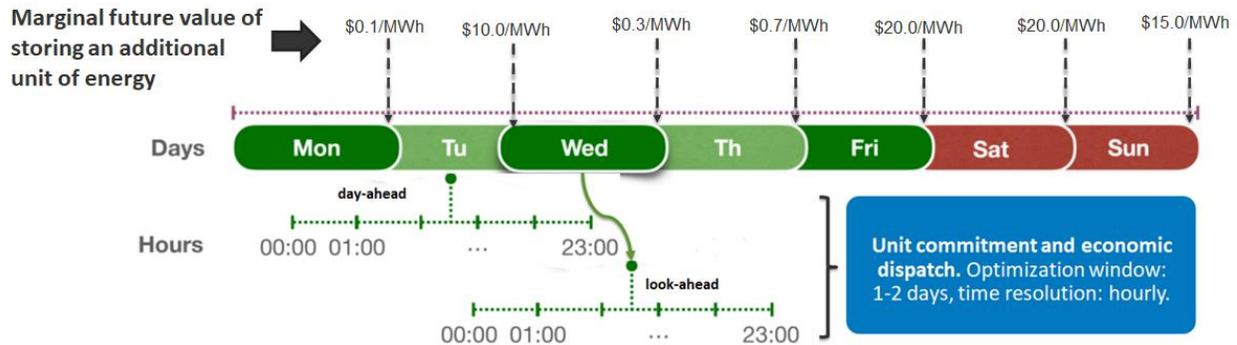

Figure 4. Stored energy value, e.g., $15.0/MWh at the end of the day, approach for modeling long-duration and seasonal energy storage.

Although this methodology is simple in theory, implementation is not usually as straightforward. The crux of this method involves assuming a value that should be applied to the stored energy. The optimal value to choose may not be easy to ascertain and is likely to vary by location, season, and other system conditions (such as prevalence of zero-marginal cost resources or natural gas prices). An incorrect choice of this value will lead to sub-optimal dispatch, either rarely dispatching stored energy or failing to see the future value of holding onto stored energy. This methodology could also lead to "over-dispatch" where the energy storage device is chasing both high and low prices. This type of dispatch strategy could also lead to additional strain on the energy storage device while not actually providing additional benefit.

## 2.4 Variable time step

The variable time step approach [19] uses a variable time step to reduce the complexity of the unit commitment and economic dispatch problems by aggregating some consecutive hours, while keeping the associated chronology between time steps. To this end, critical periods—periods in which hourly temporal resolution is required (e.g., peak net load hour, 4-hours peak period without the peak net load, etc.) are identified *a priori* and the remaining hours are aggregated based on a given daily sub-sampling strategy, as shown in Figure 5. This variable time step approach allows for the optimization of unit commitment and economic dispatch decisions over longer optimization windows, e.g., 1- or multi-year timeframes, using chronological time steps. Thus, multiscale energy storage deployment and operation can be integrated into the economic and unit commitment dispatch formulations. However, it is difficult to identify critical periods *ex ante*. For example, critical periods could depend on the deployment of energy storage and other flexibility options. Moreover, this approach assumes that system operators will dispatch the power system based on the unit commitment and economic dispatch optimization for the entire year simultaneously, assuming perfect foresight of the power system generation capacity and needs for the entire year. However, this assumption contrasts with the current electricity market settings, e.g., dispatch based on the optimal unit commitment and economic dispatch for 1 day-ahead. Moreover, the selected critical periods could depend on the load, wind generation, and solar PV generation time series or input data, which could vary across scenarios used in production cost modeling studies. Thus, the implementation of the variable time step approach can be a challenge, particularly for production cost modeling studies that involve sensitivities around load, wind generation, and solar PV generation data. Note that the variable time step approach is a specific time-series aggregation approach that preserves the chronology of the variable time steps. A more comprehensive and generic overview of time-series aggregation approaches for modeling of energy systems is presented in the literature [60].

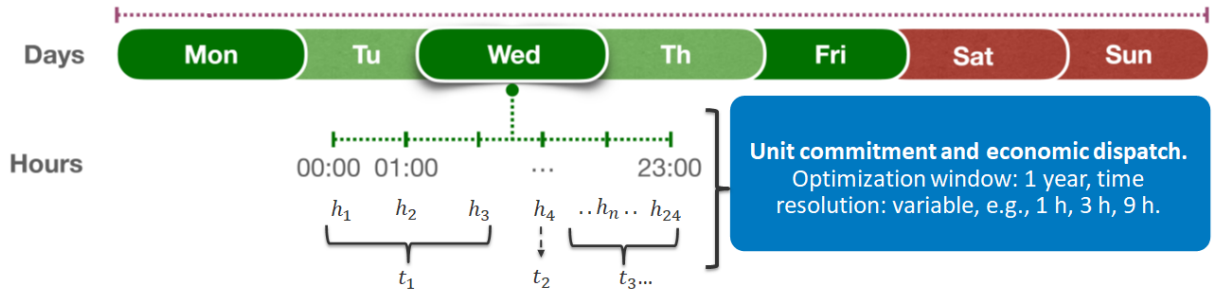

Figure 5. Variable time-step approach for modeling long-duration and seasonal energy storage (Based on De Guibert et al. 2020[19]).

This approach has been applied to investment and dispatch power system models for France and Germany [19]. First, seven critical periods were defined based on the peak net load. The durations of these critical periods were 1 h, 4 h, 6 h, 12 h, 24 h, 48 h, and 96 h. The results in terms of system cost, renewable generation, storage operation, and computational time, were compared with those of the original model formulation (e.g., chronological hourly resolution). In summary, the test showed that the variable time step approach allows for a computational time reduction of up to 60-fold, while the error in total system cost, renewable generation, and storage deployment and generation is equal or lower than 1%, 2%, and 20%, respectively. However, 1% total system cost or production cost could be equivalent to billions of dollars for large-scale power systems. Note that these results did not consider power imports and exports. If that is the case, the definition of critical periods should be based on net load, imports, and exports, which could represent a challenge given the fact that power imports/exports depend on the deployment and operation of renewable power and storage technologies.

## 2.5 Uncertainty quantification

Uncertainty in model parameters, such as differing input/output efficiencies or self-discharge rate (parametric uncertainty), and in model framework and optimization technique (structural uncertainty) can significantly impact the optimal operation of energy storage technologies [27,42]. Neglecting parametric uncertainties or the gradual revelation of parametric uncertainty could result in underestimation of power system flexibility needs and, therefore, underestimations in the flexibility value of energy storage technologies [20]. Thus, the quantification of uncertainties is key for the comprehensive modeling of energy storage. To this end, Monte Carlo simulation or sensitivity analysis can be used to quantify the impacts of parametric uncertainties on energy storage deployment and operation. On the other hand, the quantification of structural uncertainties could be addressed by comparing different mathematical formulations and evaluating the effects of different model formulations on the deployment and operation of energy storage devices [27]. Stochastic, robust, or chance-constraint optimization are more systematic approaches for handling uncertainty [20] that have been applied to energy storage optimization problems. However, these methods often face scalability and tractability issues, particularly for large-scale power system planning and PCMs.

Perhaps the most important uncertainty in long-duration storage modeling is the operational uncertainty associated with load and renewable energy resources[61]. Even in the current power system operational paradigm, this is not handled in a manner that is most beneficial for long-duration storage. This is because current numerical weather prediction models, which serve as the basis for operational load, wind, and solar forecasts, do not extend beyond approximately ten days. Even this amount of foresight is not commonly used in operations, with most power system-relevant forecasts only being endogenously used in operations out to approximately 60 hours-ahead of the operating hour. Because the proper handling of this uncertainty in long-duration storage modeling is such an open question, and because of the challenges associated with creating realistic synthetic forecasts[62], especially at time frames far ahead of current practice, we have opted in this work to utilize the assumption of perfect foresight in the modeling work that follows, with a sensitivity around VRE generation uncertainty for the look-ahead horizon. The ideal case discussed below utilizes perfect foresight over very long look-ahead windows that will never happen in practice, but provides a conservative estimate of how long-duration storage can mitigate power system load and renewable resource *variability* issues. The other modeling cases utilize the perfect foresight assumption for look-ahead windows that are in line with current operational forecasting time frames, and thus estimate how long-duration storage can mitigate variability issues within those time frames but neglect issues associated with *uncertainty* outside of those look-ahead windows. In all of these cases one could expect the realized system costs to be higher than modeled due to the uncertainty in load and renewable energy generation, though all provide a glimpse at how operational practices will influence the value of LDES, even without a full endogenous consideration of uncertainty at all timescales.

# 3 Testing the performance and scalability of long-duration modeling approaches

This section focuses on assessment of the performance of different proposed dispatch approaches for long-duration storage, based on two test power systems and considering both solar PV- and wind-driven high renewable energy shares. The performance is quantified in terms of economic, operational, and computational metrics.

## 3.1 Production cost and energy storage modeling

This section describes simplified formulations of the production cost and energy storage models that are typically implemented in commercially available power system dispatch tools. The PCM formulation consists of the minimization of the total operation (production) costs ($ProdCost$), including variable operations and maintenance, fuel, and start up and shutdown costs. Additionally, the PCM includes both power system constraints (system demand balance and power balance for each bus) and asset constraints (generators' capacity, ramping, and startup and shutdown constraints). The unit commitment and economic dispatch problems are typically formulated as mixed integer linear programming models based on the direct current optimal power flow problem. Details on the formulation of the unit commitment and economic dispatch problems have been well documented [63–65]. As a reference, a generic formulation of a PCM is presented below.

The objective function of the PCM is given by,

$$ProdCost = \sum_{t=1}^{NT} \sum_{i=1}^{NU} C_{i,t}^{fuel} \cdot p_{i,t} + C_{i,t}^{start} \cdot x_{i,t}^{start} + C_{i,t}^{stop} \cdot x_{i,t}^{stop}. \qquad 1$$

Where, the parameters $C_{i,t}^{fuel}, C_{i,t}^{start}, C_{i,t}^{stop}$ represent the fuel cost, startup cost, and shutdown cost of thermal generator '$i$' at time '$t$', respectively. The variables in the objective function include thermal generation ($p_{i,t}$), and binary variables representing the starting and stopping of thermal generators ($x_{i,t}^{start}, x_{i,t}^{stop}$). The objective function is optimized for the following set of constraints.

$$\sum_{i=1}^{NU} p_{i,t} + \sum_{s=1}^{NS} p_{s,t}^d = \sum_{j=1}^{J} d_{j,t} + \sum_{s=1}^{NS} p_{s,t}^c \qquad \forall t \qquad 2$$

$$P_{i,t}^{min} \cdot x_{i,t} \leq p_{i,t} + \sum_{r} r_{r,i,t} \leq P_{i,t}^{max} \cdot x_{i,t} \qquad \forall i,t \qquad 3$$

$$R_{i,t}^{down} \leq p_{i,t} - p_{i,t-1} \leq R_{i,t}^{up} \qquad \forall i,t \qquad 4$$

$$x_{i,t-1} - x_{i,t} + x_{i,t}^{start} - x_{i,t}^{stop} = 0 \qquad \forall i,t \qquad 5$$

$$\sum_{i} r_{r,i,t} + \sum_{s} r_{r,s,t} \geq RR_{r,t} \qquad \forall r,t \qquad 6$$

$$\sum_{j \in B[k]}^{J} f_{t,j,k} = \sum_{i \in Q[k]}^{NU} p_{i,t} - d_{k,t} + \sum_{s \in U[k]}^{NS} p_{s,t}^d - \sum_{s \in U[k]}^{NS} p_{s,t}^c \qquad \forall k,t \qquad 7$$

$$F_{j,k}^{min} \leq f_{t,j,k} \leq F_{j,k}^{max} \qquad \forall t,j,k \qquad 8$$

$$f_{t,j,k} = B_{j,k} \cdot (\theta_{j,t} - \theta_{k,t}) \qquad \forall t,j,k \qquad 9$$

$$\theta^{min} \leq \theta_{k,t} \leq \theta^{max} \qquad \forall k,t \qquad 10$$

Equation (2) represents the total energy demand-supply balance, where for each time interval '$t$', the sum of generation must be equal to the sum of energy demand ($d_{k,t}$) on all nodes '$k$', including storage losses. Equations (3)–(5) are constraints associated with the operational constraints for generators, e.g., thermal generators, in the electricity grid. Equation (3) defines the operating envelope for the committed generators by specifying lower and upper bounds on the generation ($P_{i,t}^{min}$ and $P_{i,t}^{max}$, respectively). Equation (4) restricts the time rate of generation change within ramping capabilities ($R_{i,t}^{down}$, $R_{i,t}^{up}$). Equation (5) tracks the generator unit commitment status ($x_{i,t}$) with startup ($x_{i,t}^{start}$) and shutdown ($x_{i,t}^{stop}$) binary variables. Equation (6) ensures that total of reserves provided by generators ($\sum_r r_{r,i,t}$) and storage devices ($\sum_s r_{r,s,t}$) is at least equal to the reserve requirement at that time ($RR_{r,t}$). Note that all reserves are assumed to be of "raise" type in our model. Equations (7)–(10) represent simplified transmission constraints. The nodal power balance (Eq. 10) keeps track of the power flow along each line (connecting nodes '$j$' and '$k$' given by '$f_{t,j,k}$'). $B$, $Q$, and $U$ represent sets of buses, generators, and storage units, respectively. Equation (8) provides minimum and maximum power flow limits for transmission lines. Equation (9) evaluates bus voltage angles with the help of line susceptance ($B_{j,k}$) and Equation (10) specifies minimum and maximum limits for the bus voltage angle '$\theta$'.

The storage model formulation consists of a set of constraints for each storage device in the power system, e.g., , $s \in NS$, as shown in equations (11)-(16), including the constraints for charging ($p_{s,t}^c$) and discharging ($p_{s,t}^d$) power capacities, as expressed by equations (11) and (12), respectively. Index $t$ denotes hours in the analysis period, e.g., 8,760 hours represented by the set $NT$ ($t \in NT$). A binary variable ($x_{s,t}^c$) tracks the operating mode of the storage device, e.g., $x_{s,t}^c = 1$ if the storage device is in charging mode and $x_{s,t}^c = 0$ otherwise. Parameters $PC_s^{max}$ and $PD_s^{max}$ denote the maximum charging and discharging power capacities for storage device s, respectively. The energy balance for each storage device is expressed as function of the SOC ($SoC_{s,t}$), as expressed by equation (13), whereas equality constraint (14) is implemented to endogenously optimize the initial SOC of each storage device. Parameters $\eta_s^{sd}$, $\eta_s^c$, and $\eta_s^d$ denote the self-discharge rate, the charging efficiency, and the discharging efficiency, respectively. Equation (15) imposes energy capacity limits. Parameters $SoC_s^{min}$ (minimum energy capacity) and $SoC_s^{max}$ (maximum energy capacity), denote the minimum and maximum SOC for each storage device, respectively. Finally, equation (16) accounts for the raise reserve services provided by energy storage devices, by either increasing the power discharge or decreasing the charging power at a given time.

$$0 \leq p_{s,t}^c \leq PC_s^{max} \cdot x_{s,t}^c \quad \forall\, t \in NT, s \in NS \qquad (11)$$

$$0 \leq p_{s,t}^d \leq PD_s^{max} \cdot (1 - x_{s,t}^c) \quad \forall\, t \in NT, s \in NS \qquad (12)$$

$$SoC_{s,t} = (1 - \eta_s^{sd}) \cdot SoC_{s,t-1} + p_{s,t}^c/\eta_s^c - \eta_s^d \cdot p_{s,t}^d \quad \forall\, t \in NT, t > 1, s \in NS \qquad (13)$$

$$SoC_{s,t=1} = (1 - \eta_s^{sd}) \cdot SoC_{s,t=|NT|} + p_{s,t=1}^c/\eta_s^c - \eta_s^d \cdot p_{s,t=1}^d \quad \forall\, s \in NS \qquad (14)$$

$$SoC_s^{min} \leq SoC_{s,t} \leq SoC_s^{max} \quad \forall\, t \in NT, s \in NS \qquad (15)$$

$$p_{s,t}^d + \sum_r r_{r,s,t} \leq PD_s^{max} + p_{s,t}^c \quad \forall\, t \in NT, s \in NS \qquad (16)$$

Note that the simplified storage model represented by equations (11)-(16) is generic and is not intended to represent any specific technology. Thus, this model does not contemplate technology-specific constraints or modeling issues, including degradation, e.g., for electrochemical energy storage, effects of charging and discharging rates on efficiencies, dependency of power on the SOC, parasitic losses, or restricting number of storage cycles among others [66–69]. But the above constraints can be integrated in the existing PCM for the technology-specific storage dispatch analysis. The simplified formulation represented by equations (1) – (16) also does not include modeling details, such as generator forced outages and generator up/down time constraints. However, this basic formulation can be modified to incorporate these thermal generator technical details in the PCM.

## 3.2 Test power systems, scenarios, and assessed modeling approaches.

Although we have thus far identified several methods for dispatching LDES, the appropriate approach might vary based on the questions to be answered. To better examine the tradeoffs between the various methodologies we perform simulations on two modified test power systems: the PJM 5-bus [70] and the Reliability Test System Grid Modernization Lab Consortium (RTS-GMLC) test system [71,72]. The PJM 5-bus test system, as configured for this work and shown in Figure 6, has 5 buses, 3 load centers, 6 transmission lines, 8 generators (including 2 solar PV facilities and 1 wind facility, which are not shown in Figure 6), 1 short-duration storage device (with 80% roundtrip efficiency and 4 hours of duration), and 1 long-duration storage device (with 65% roundtrip efficiency and 10 hours of duration). The RTS-GMLC test system, which is an updated version of the RTS-96 test system shown in Figure 7, has 74 buses, 14 load centers, 121 transmission lines, 159 generators (including 57 solar PV facilities and 4 wind facilities, which are not shown in Figure **7**), 1 short-duration storage device (with 80% roundtrip efficiency and 4 hours of duration), and 1 long-duration storage device (with 65% roundtrip efficiency and 15 hours of duration). As discussed previously, the integration of zero-marginal-cost VRE resources provides opportunities and challenges for long-duration storage dispatch, so we increased wind, solar PV, and energy storage resources to create systems that have a 70% - 90% VRE annual energy mix based on the annual load. Single year synchronous weather data was used for wind and solar PV profiles. The installed generation capacity is shown in Figure 8 for each system and for two cases: one dominated by solar PV and one by wind power. Moreover, regulation up and spinning contingency reserve constraints (hourly) were included for all tests systems and cases. For the PJM 5-bus test system regulation up reserves are 5% of load and spinning up reserves are 5% of load plus 10% of wind availability (forecast) plus 4% of solar PV availability (forecast). Reserve constraints for the RTS-GMLC tests system are based on the time series for each reserve product provided by the previous study[72].

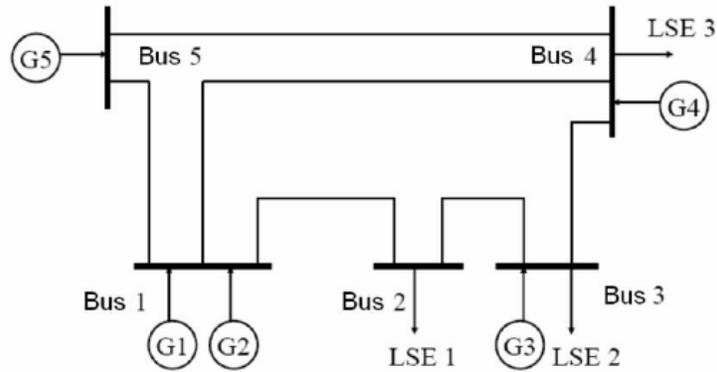

Figure 6. PJM 5-bus test system[73].

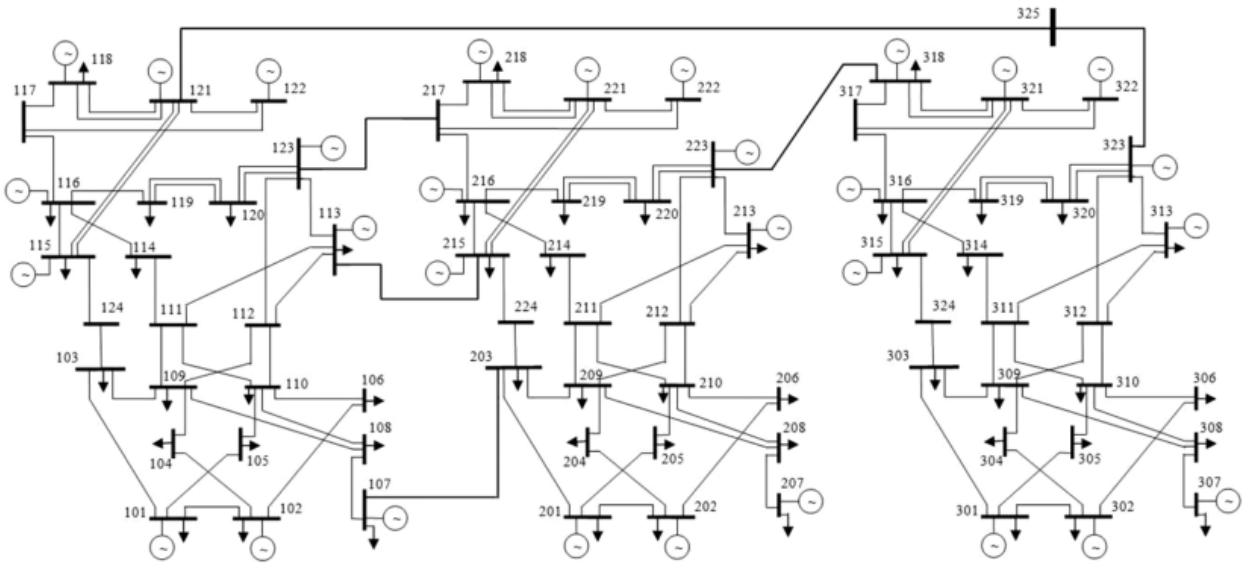

Figure 7. RTS-96 test system[74].

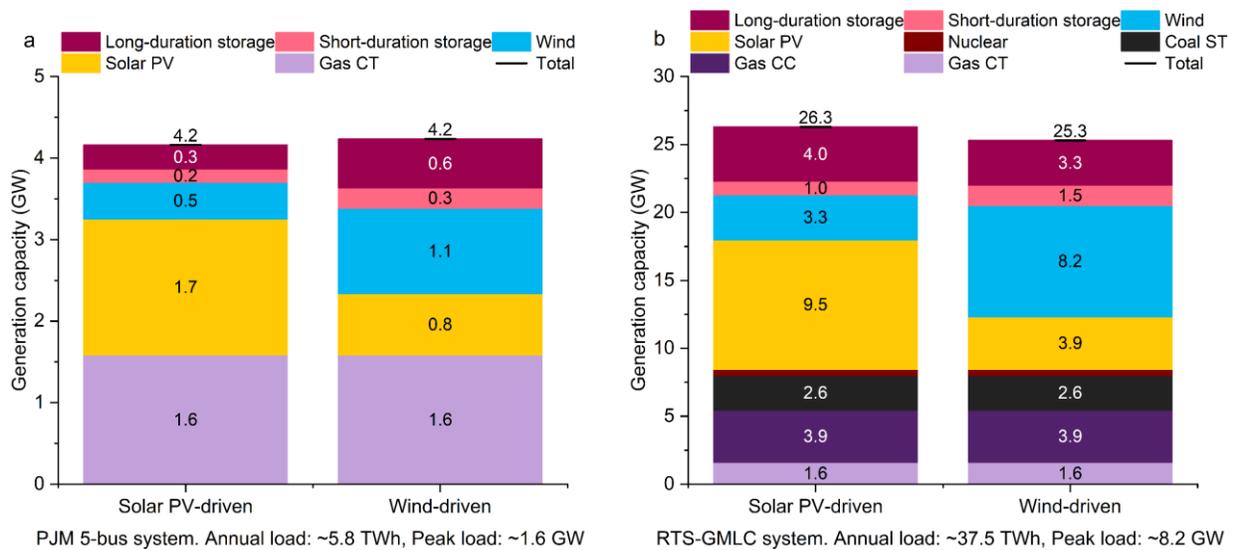

Figure 8. Capacity mixes for the modified PJM 5- bus (a) and Reliability Test System Grid Modernization Lab Consortium (RTS-GMLC) (b) solar PV- and wind-driven test power systems.

These test systems were modeled in the open-source platform Sienna (https://github.com/NREL-Sienna), which uses the Julia v1.7 programming language and the Xpress v11 solver [75]. NREL's high-performance computing system was used for computational analysis. The computing system features Dual Intel Xeon Gold Skylake 6154 (3.0 GHz, 18-core) processors and 192 GB RAM that runs on the Linux operating system. The solver's relative optimality gap was fixed at 0.001% for all of the different modeling approaches tested.

The following storage modeling approaches were tested to evaluate their capabilities and limitations: (i) extended optimization horizon, (ii) end volume targets, and (iii) stored energy value. Moreover, these methods were compared with the ideal solution approach, defined as a 1-year optimization horizon with hourly resolution with perfect foresight. The solution of the linear programing (LP) relaxation of the ideal case was also included as a reference, i.e., a lower bound for the production cost (upper bound for the production cost reductions), for case where the ideal solution is not available. However, the LP solution is likely infeasible/unrealistic because of the relaxation of unit commitment constraints. Thus, the LP solution should not be used to assess the performance of any dispatch approach. Additionally, the variable time step approach was excluded because the implementation of this approach could depend on the input data (load time series, wind generation, solar PV generation, and wind to solar PV generation mix), which makes this approach less attractive from a practical viewpoint, as described previously in Section 2.4.

Table 1 shows a description and notation of the long-duration modeling approaches examined here. Note that different settings or parameters were evaluated to identify the most promising implementations for each method, which are summarized in Table 1. For example, for the extended optimization horizon or window of foresight, we tested 3 days, 1 week, and 1 month of look-ahead. Note that for 1 week and 1 month of look-ahead, a time limit of 1000 seconds was enforced on the solver for each unit commitment and economic dispatch sub-problem. For example, in case of a 1 week look-ahead problem, we solve 359 sub-problems (359 days = 366 days of planning horizon – 7 days of look-ahead). For each sub-problem, a time limit of 1000 seconds is imposed. For an ideal case, a time limit of 1 day was enforced to solve the unit commitment and economic dispatch problem for an entire year. Following are the optimality gaps achieved for each ideal case: 0.029% for PJM 5-bus wind-driven, 0.001% for PJM 5-bus solar-PV driven, 0.25% for RTS-GMLC solar PV-driven, and 0.18% for RTS-GMLC wind-driven. For the end volume targets method, we tested a 50% SOC target at the end of the 24 hours look-ahead horizon, and MT scheduling-based SOC targets at the end of the 24-hour look-ahead horizon. Note that the MT scheduling model includes the following characteristics: 1 week optimization horizon (1 week-ahead), 1 week look-ahead, hourly resolution for both optimization and look-ahead windows, operating reserves, and no commitment constraints. Finally, for the energy value approach, we tested two constant future energy values (e.g., $0.1/MWh and $0.25/MWh). We also performed preliminary testing of variable energy values, based on net-load and variable operating cost of conventional generators, but the results were inconclusive. For instance, the performance of a given implementation of the energy value approach based on dynamic prices was not consistent across the cases tested in this study.

Table 1. Description of the long-duration modeling approaches examined.

| Method | Notation | Optimization Horizon | Look-ahead horizon | Parameter/Detail |
|---|---|---|---|---|
| Ideal | ID | 8760 h | - | - |
| Ideal LP | ID-LP | 8760 h | | Linear programming (LP) relaxation |
| Traditional | TR | 24 h | 24 h | - |
| Extended look-ahead | ELH-3d | 24 h | 72 h (3 days) | - |
| Extended look-ahead | ELH-1w | 24 h | 168 h (1 week) | - |
| Extended look-ahead | ELH-1m | 24 h | 720 h (1 month) | - |
| End volume targets | EVT-LA | 24 h | 24 h | 50% SOC @ hour 48 |
| End volume targets | EVT-LA-MT | 24 h | 24 h | MT SOC @ hour 48 |
| Energy value | EV-01 | 24 h | 24 h | $0.1/MWh |
| Energy value | EV-025 | 24 h | 24 h | $0.25/MWh |

The traditional approach of formulating a PCM involves one day of optimization horizon and one day of the look-ahead period. The objective function and constraints are formulated for 48 hours with hourly temporal resolution. For example, Equation 1 has 48 time periods representing the energy demand-supply balance for each hour in the optimization window. In the traditional dispatch approach, the unit commitment and economic dispatch decisions in the optimization horizon are reported, while decisions in the look-ahead period are discarded. The objective function in the traditional approach minimizes the production cost for 48 hours (as shown in Equation 16). Note that the ideal case is equivalent to the formulation based on Equations (1)-(16) and considering 8760 time periods simultaneously.

$$\sum_{t=1}^{48} \sum_{i=1}^{NU} C_{i,t}^{fuel} \cdot p_{i,t} + C_{i,t}^{start} \cdot x_{i,t}^{start} + C_{i,t}^{stop} \cdot x_{i,t}^{stop} \qquad 17$$

As the name suggests, the extended look-ahead method features an increase in the number of days in the look-ahead period. For example, Equation 18 shows the objective function for a 3-day look-ahead optimization problem. Note that the number of equations that need to be solved increases rapidly as the look-ahead period extends.

$$\sum_{t=1}^{96} \sum_{i=1}^{NU} C_{i,t}^{fuel} \cdot p_{i,t} + C_{i,t}^{start} \cdot x_{i,t}^{start} + C_{i,t}^{stop} \cdot x_{i,t}^{stop} \qquad 18$$

The energy value method incentivizes storage to charge more (i.e., maintain a higher SOC) by modifying the objective function of the PCM. Equation (19) features an additional term in the objective function – a product of energy value ($EV$), e.g., \$/MWh, and SOC ($SoC_{s,t}$) [MWh] that gets subtracted from the production cost. In this work, the energy value is associated with the end-of-the-day SOC (i.e., $t\dot{} = \{24, 48\}$).

$$\sum_{t=1}^{48} \sum_{i=1}^{NU} C_{i,t}^{fuel} \cdot p_{i,t} + C_{i,t}^{start} \cdot x_{i,t}^{start} + C_{i,t}^{stop} \cdot x_{i,t}^{stop} - \sum_{s=1}^{NS} \sum_{t=\{24,48\}} EV \cdot SoC_{s,t} \qquad 19$$

The energy volume target method specifies the SOC for the storage at the end of a given hour (Equation 20) as an additional constraint in the PCM.

$$SOC_{s,t=T'} = SoC_{s,t=T'}^{target} \qquad 20$$

The introduction of a constraint like Equation 20 can sometimes produce an infeasible solution due to hard equality. In such a case, it could be convenient to convert Equation 20 into a soft constraint by adding a penalty term in the objective function. However, a drawback is associated with the "softening" of the energy volume target constraint. The objective function of the PCM usually contains additional penalty terms to avoid dropped load and thermal curtailment. In such cases, the value of the penalty term needs to be adjusted carefully to avoid prioritizing the energy volume target over dropped load and curtailment.

## 3.3 Quantitative metrics for the performance of LDES dispatch approaches

This section describes the quantitative metrics used to assess the performance of the LDES dispatch methods listed in Table 1, and compare the overall system value. We define system value as comprising two parts: (i) operational value, which is the amount of improvement in total production cost (% of total production cost); and (ii) capacity credit, which is the ability of the long-duration storage device to supply energy during the 10 hours of peak net load (e.g., based on the discharged power) where net load is defined as total load minus total forecasted VRE production [76,77]. Note that one potential limitation of this standard capacity credit estimation, which corresponds to the capacity factor-based approximation of the capacity value[21], is that the stored energy in the storage device during peak net load hours is not considered. Thus, we propose a novel SOC-aware peak net load-based capacity credit estimation, which considers not only the discharged power but also the stored energy during peak net load periods.

The standard peak net load-based capacity credit, e.g., $CC_s$, for a given storage device, e.g., $s$, is calculated based on the top 10 peak net load hour, as expressed by Equation 21. Index '$t'$' denotes the top peak net load hours, e.g., $t' = 1$ represents the hour with the peak net load of the system.

$$CC_s = 100 * \frac{\sum_{t'=1}^{t'=10} p_{s,t'}^d}{10 * PD_s^{max}} \qquad \forall s \qquad 21$$

One potential limitation of the standard peak net load-based capacity credit is that it does not consider the SOC of the storage device, which provides operation flexibility to address the load and generation balance during peak net load hours. Thus, here we propose a new SOC-aware peak net load-based capacity credit, e.g., based on the top 10 peak net load hours, as expressed by Equation 22. Note that the additional term in the numerator, e.g., $Min\{PD_s^{max} - p_{s,t'}^d, \eta_s^d. SOC_{s,t'}\}$, represents the additional power that could be discharged during peak net load hour $t'$ from the storage device using the stored energy. Note that this novel capacity credit metric has two assumptions or caveats: (i) for a given storage device, dispatching 1 MW of power during a peak net load hour, is equivalent to having a SOC ≥ 2 MWh (assuming a discharge efficiency of 50%) and no power output from the storage device during that peak net load hour, and (ii) it is assumed that there is no chronology between the peak net load hours, e.g., using the stored energy at hour $t' = 1$ will not affect the availability of stored energy during hours $t' = 2, t' = 3, t' = 4$, etc.

$$CC_s = 100 * \frac{\sum_{t'=1}^{t'=10}(p_{s,t'}^d + Min\{PD_s^{max} - p_{s,t'}^d, \eta_s^d.SOC_{s,t'}\})}{10 * PD_s^{max}} \qquad \forall s \qquad 22$$

We also evaluated computational and LDES operational aspects, including: (i) required CPU time, (ii) peak memory usage, (iii) LDES equivalent annual cycles, and (iv) VRE curtailment. Note that unserved load was insignificant in all of the cases, thus, it was not included as a metric for the assessment of the LDES dispatch approaches (however, this metric could be useful for reliability aspects of the system under extreme events or outages).

# 4 Quantitative results

This section summarizes the results for the tested LDES dispatch approaches, based on the metrics defined in the previous section. First, based on the assumption of no transmission constraints, we evaluated the potential system value of modeling long-duration energy storage based on the traditional (24 hours optimization window + 24 hours look-ahead horizon) and the ideal (8760 hours optimization window with no look-ahead) methods. Note that the ideal case assumes perfect foresight of VRE generation, load, and storage needs for the entire year and, therefore, does not provide the desired level of realism but is a useful basis for comparison. Then, we compare the performance of each dispatch method in terms of system value, (operational value and capacity value), and computational requirements (CPU time and peak memory usage). We then illustrate how including transmission constraints and VRE uncertainty could impact the performance of the most promising LDES dispatch approach. Finally, we provide future research directions based on the insights from the evaluated LDES modeling approaches and test power systems.

## 4.1 The potential system value of modeling long-duration energy storage

First, we evaluate the potential value of LDES by comparing the ideal (ID) case and the traditional (TR) cases, assuming perfect foresight. Overall, the ideal LDES dispatch has the potential to reduce power system production cost by 4%-14%, relative to the traditional method (shown in Figure 9). These cost reductions are mostly driven by fuel cost savings, which indicate a more effective use of LDES to displace thermal generators, including peaking conventional generators (e.g., gas power plants). Indeed, it was found that the ideal LDES dispatch approach increases the LDES standard capacity credit (based on the top 10 peak net load hours) from 75% to 97% and from 58% to 72% for the PJM 5-bus and RTS-GMLC solar PV-driven test power systems, respectively. The results for wind-driven systems are similar, i.e., an increase of the standard capacity credit from 42% to 68% and from 52% to 86% for the PJM 5-bus and RTS-GMLC test power systems, respectively.

Note that regardless of the test system and the energy mix, the SOC-aware peak net load-based capacity credit for LDES was always 100% for the ideal dispatch approach. In contrast, regardless of the test system and energy mix, the ideal LDES dispatch approach reduces the standard capacity credit of short-duration energy storage. For example, the standard capacity credit of short-duration energy storage is reduced from 75% to 73% and from 79% to 67% for the PJM 5-bus and RTS-GMLC solar PV-driven test power systems, respectively. Similarly, these standard capacity credit reductions are from 68% to 59% and from 64% to 28% for the PJM 5-bus and RTS-GMLC test power systems, respectively.

Note also that regardless of the test system and the energy mix, the SOC-aware peak net load-based capacity credit for short-duration was greater than or equal to 73% for the ideal dispatch approach. However, regardless of the test system and energy mix, the ideal LDES dispatch approach increases the standard capacity credit of total energy storage capacity (combined short-duration and LDES) (e.g., an increase between 8.8% and 15.7% on the standard capacity credit of the total energy storage capacity). Moreover, for the traditional dispatch approach, LDES provide between 92% and 100% of the total required reserves during the top 10 net load hours, depending on the test system and energy mix. The LDES reserve provision is between 90% and 100% for the ideal dispatch approach. Additionally, the ideal dispatch allows for a better utilization of the LDES (e.g., the equivalent annual discharge cycles increase by up to 37% when compared with the traditional dispatch). Note that these results are based on the assumption of perfect foresight of VRE generation, load, and storage needs for the traditional and ideal dispatch approaches. In practice, these parametric uncertainties are gradually revealed and therefore the assumption of perfect foresight could result in underestimation of power system flexibility needs, which could impact the estimated value of modeling LDES.

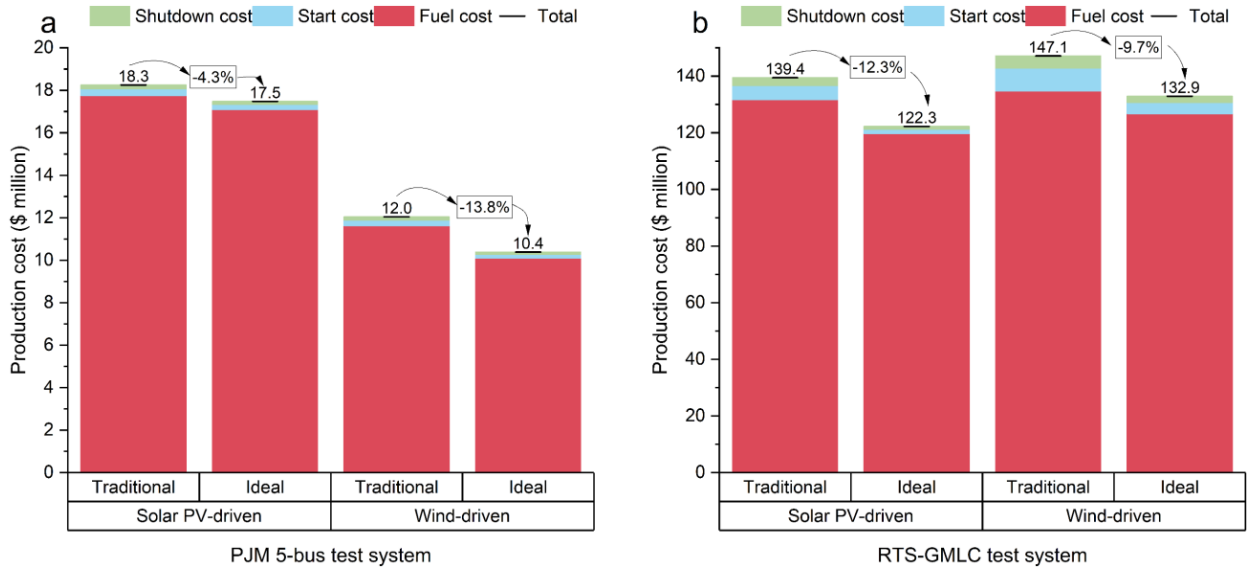

Figure 9. Production cost breakdown for PJM 5- bus (a) and Reliability Test System Grid Modernization Lab Consortium (RTS-GMLC) (b) solar PV- and wind-driven test power systems.

Depending on the power system and the VRE mix, the ideal LDES dispatch could increase VRE contribution by 1.2% - 5% while decreasing VRE curtailment by 1.3% - 3.3% (using otherwise curtailed energy), as illustrated in Figure 10. This reduction in VRE curtailment decreases the use of fuel for power generation and associated $CO_2$ emissions, which explains the reduction in production cost observed in Figure 9. Thus, the ideal LDES dispatch has the potential to facilitate the integration of VRE by allowing a more effective utilization of LDES devices by allowing for more inter-day VRE shifting, while reducing the need for power generation from thermal power plants. Note that there is a significant VRE curtailment for the wind-driven cases, this could be because the storage power capacities and durations assumed in this study were not optimized for the cases evaluated in this study. However, the high VRE curtailment is consistent across the results from both traditional and ideal dispatch approaches.

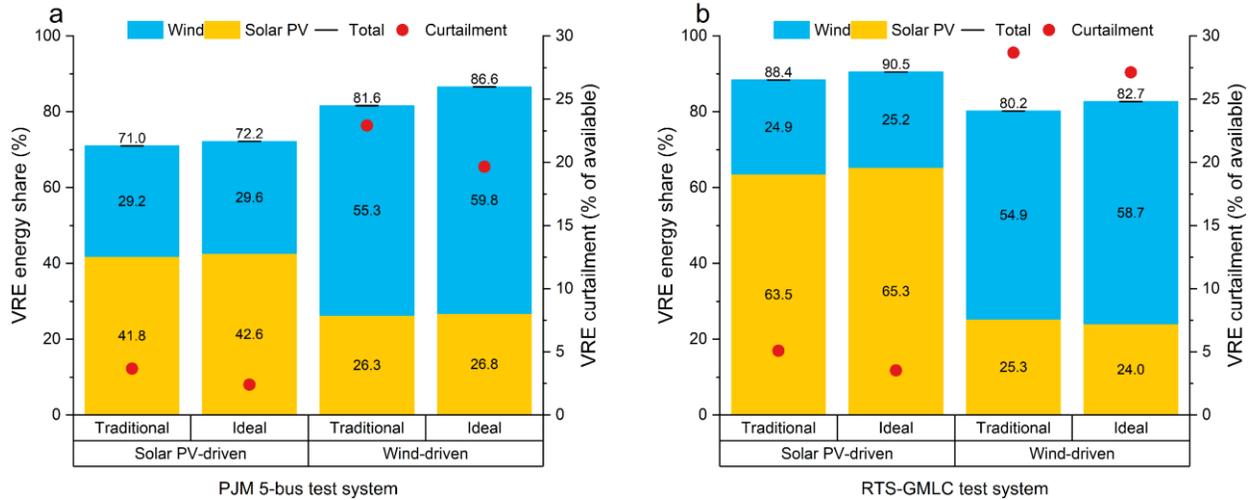

Figure 10. Variable renewable energy (VRE) integration and curtailment for PJM 5- bus (a) and Reliability Test System Grid Modernization Lab Consortium (RTS-GMLC) (b) solar PV- and wind-driven test power systems.

Overall, we find marked reductions in production cost from the ideal LDES dispatch case compared to the traditional LDES dispatch case. However, as the ideal LDES dispatch is computationally infeasible for real-world systems, we use this as an upper bound for the possible value of LDES and can serve as a benchmark for comparing other dispatch routes, as discussed below.

## 4.2 Trade-off between improved representation and computational complexity in modeling long-duration energy storage

Using the ideal and traditional dispatch methods as benchmarks, this section considers the performance of other dispatch methods. Regardless of the test power system and the VRE mix, the extended optimization horizon or window of foresight and the end volume targets dispatch approaches always reduce the production cost of the power grid when compared with the traditional dispatch approach (e.g., 1 day-ahead plus 1 day look-ahead), as illustrated in Figure 11. For instance, using the traditional dispatch approach as a baseline, depending on the power system and the VRE mix, extending the look-ahead horizon could reduce production cost between 3.5% and 13.8%, while the end volume target dispatch approach could reduce production cost between 0.1% and 12.8%. Note that adding 1 week of look-ahead horizon (extended look-ahead ELH-1w dispatch approach) achieve basically the same production cost reduction as the ideal case, with significant computational savings (see Figure 11 and Figure 13). Note that storage devices with longer duration could require longer look-ahead horizons. On the other hand, the energy value dispatch approach could increase or decrease the production cost depending on the system and the VRE mix (e.g., the long-duration storage tends to be underutilized and thus results in a worse performance). Both the extended window of foresight and the end volume targets dispatch approaches increase the standard capacity credit of LDES regardless of the power system and the VRE mix, see Figure 12. This is also true for the SOC-aware peak net load-based capacity credit. On the other hand, the standard capacity credit of short-duration decreases or increases depending on the power system and the energy mix, but the standard capacity credit of total energy storage capacity (combined short-duration and LDES) tends to increase. Additionally, for the extended window of foresight dispatch approaches, LDES provide between 18% and 100% of the total required reserves during the top 10 net load hours, depending on the test system and energy mix. The LDES reserve provision is between 18.9% and 100% for the end volume targets dispatch approaches. As with the production cost, the energy value approach could increase or reduce the standard capacity credit of LDES depending on the power system and the VRE mix. Therefore, from the system value standpoint, the extended optimization horizon and the end volume targets dispatch approaches can improve the representation of LDES in power system dispatch models. On the other hand, the energy value approach requires more tuning to find the proper parameters that allow for a better dispatch of LDES.

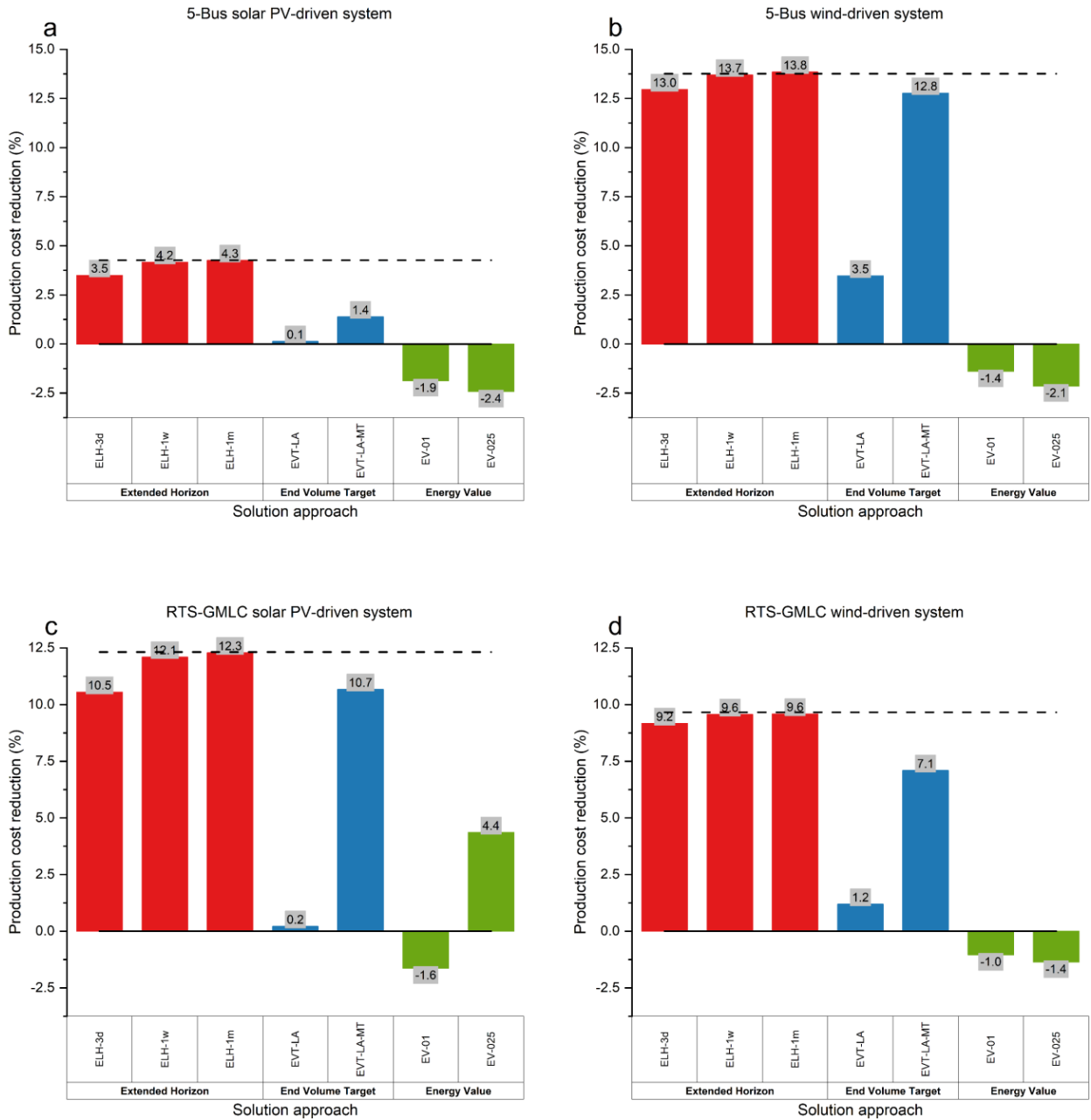

Figure 11. Production cost reduction (based on the traditional dispatch case) for PJM 5- bus (e.g., a and b), and Reliability Test System Grid Modernization Lab Consortium (RTS-GMLC) (e.g., c and d) solar PV- and wind-driven test power systems.

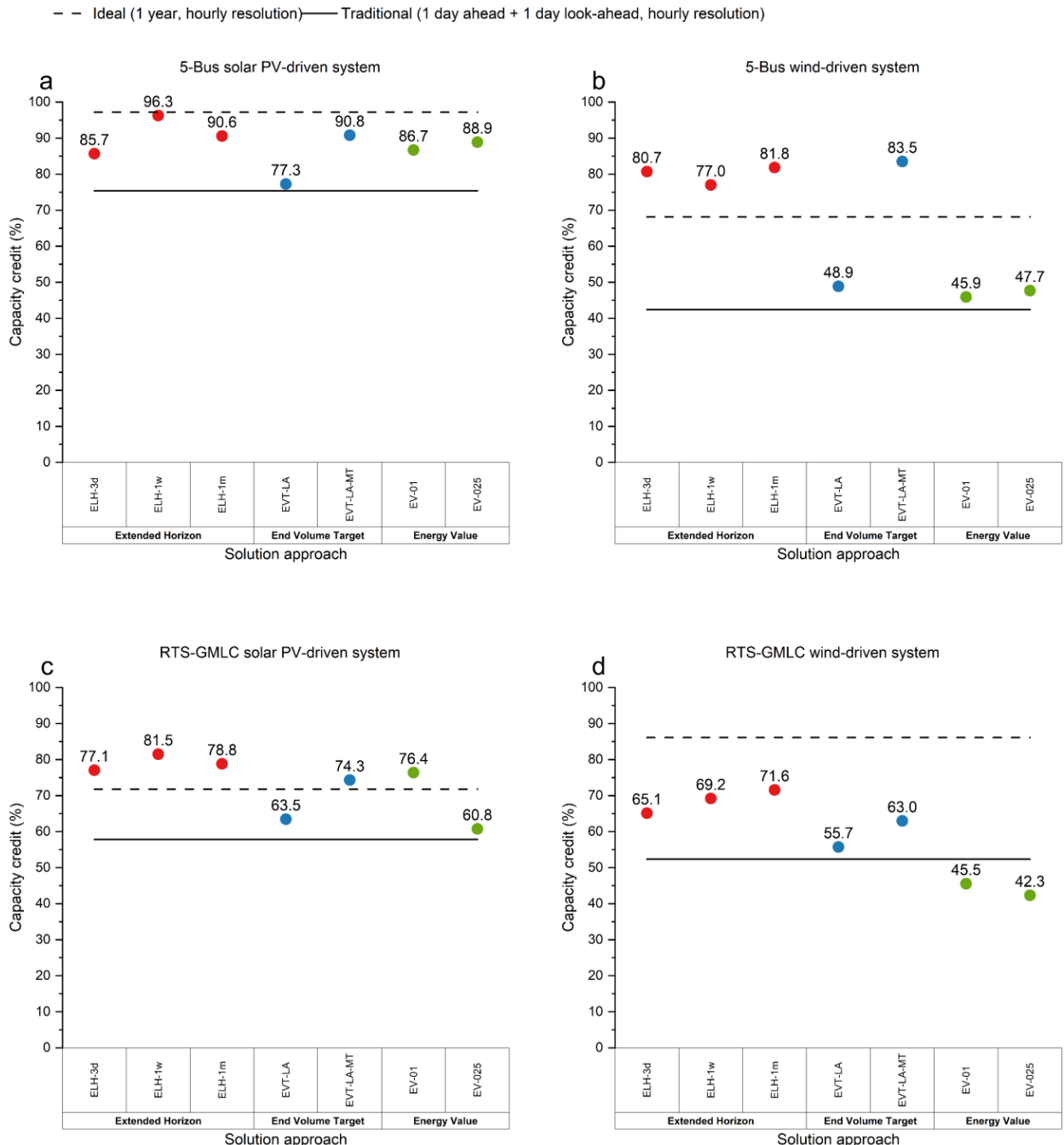

Figure 12. Standard capacity credit (based on top 10 peak net load hours) for PJM 5- bus (e.g., a and b), and Reliability Test System Grid Modernization Lab Consortium (RTS-GMLC) (e.g., c and d), solar PV- and wind-driven test power systems.

Another important aspect when evaluating LDES dispatch approaches is the scalability of the scheduling approaches measured in terms of computational requirements (e.g., CPU time and peak memory usage). It was observed that, for the extended optimization horizon, the CPU time increases exponentially with the length of the look-ahead horizon (particularly for the RTS-GMLC tests power system), as shown in Figure 13. This could limit the scalability of the extended optimization horizon or window of foresight, particularly for real world applications involving large-scale power systems with thousands of power plants, thousands of transmission lines, and thousands of buses.

In contrast, the CPU time for the end volume targets and the energy value approaches is similar to that of the traditional approach. Similar findings were observed when the peak memory usage metric was used to assess the performance for the LDES dispatch approaches. Thus, the end volume target and the energy value approaches do not significantly increase the computational requirements of PCMs. However, for the MT scheduling-based end volume target method (EVT-LA-MT), the MT model can overestimate the flexibility provided by thermal fleet as it is ignoring commitment constraints and start-up and shutdown costs, which could limit the performance of this approach in terms of improved system value for power systems with moderately-sized thermal fleets. Similarly, the use of a flat SOC target (EVT-LA method) can sometimes over or underestimate the required amount of energy going into the next day, which limits the performance of the EVT method in terms of improved system value.

On the other hand, there is a trade-off between the improved representation and the computational requirements associated with the extended optimization horizon (ELH-3d, ELH-1w, and ELH-1m approaches). Overall, this set of tests has highlighted that each of these LDES dispatch formulations have certain strengths and limitations that users should be aware of before making the choice to utilize them. For instance, extended horizon is probably a very effective tool to approximate the ideal LDES operation but suffers from scaling issues on larger systems which could be addressed by temporal decomposition methods. Energy value is not as computationally intensive to solve, although it requires some tuning of the energy value parameter. Using a static value, however, can lead the model to over- or under-estimate the value of stored energy. This shortcoming can be addressed by using a time varying energy value, but it doesn't address the need to come up with a more robust methodology to determine the energy value for a given system. One method that does strike a balance between approximating the value of LDES and computational scaling is EVT-LA-MT but the end volume targets (energy targets) are only as good as the representation of system operations in the mid-term planning model.

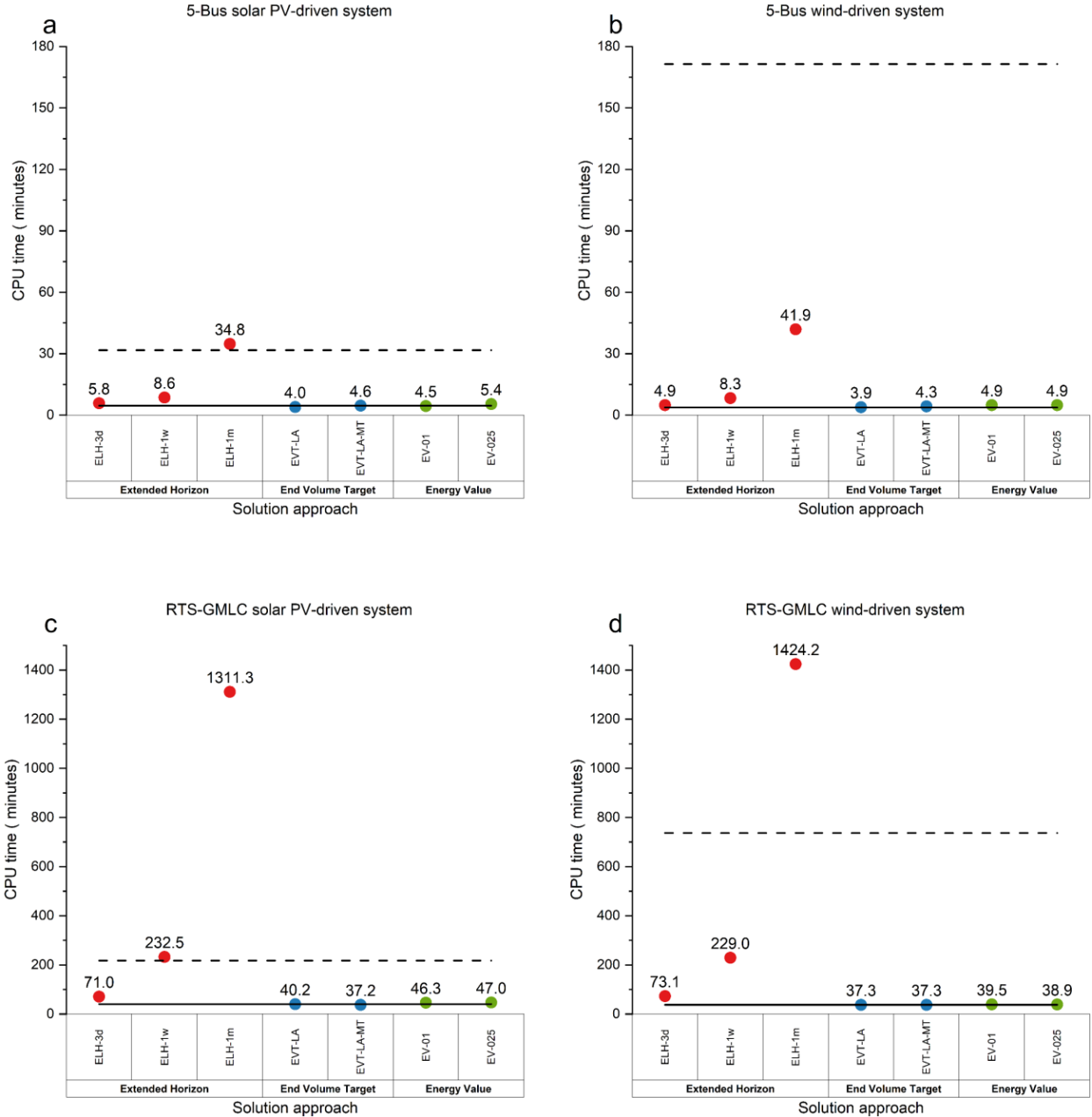

Figure 13. Central processing unit (CPU) time PJM 5- bus (e.g., a and b), and Reliability Test System Grid Modernization Lab Consortium (RTS-GMLC) (e.g., c and d), solar PV- and wind-driven test power systems.

## 4.3 Performance of long-duration energy storage dispatch approaches with transmission constraints

To test the performance of the LDES dispatch approaches for transmission-constrained power systems, we used the nodal representation of the RTC-GMLC power system to assess the LDES dispatch approaches: (i) extended optimization horizon, (ii) end volume targets, and (iii) stored energy value dispatch approaches for both solar PV- and wind-driven cases. The short-duration storage, long-duration storage, and VRE capacities were allocated (e.g., based on the load) to nodes with existing (installed) capacity of thermal generation and increasing transmission capacity by two-folds. As our work focused on the assessment of LDES dispatch approaches, the nodal allocation of storage and VRE capacity is not optimized but hopefully it is useful to assess the performance of the evaluated LDES dispatch approaches (e.g., the scalability). The results in terms of production cost and standard capacity credit are summarized in Figure 14.

Note that the ideal case (8,760 optimization window with no look-ahead), the 1-week extended window of foresight (ELH-1w), and 1 month extended window of foresight (ELH-1m) runs did not finish after 2 days and were not included in the results for the transmission-constrained cases. Indeed, these cases reinforce the scalability issues of the extended optimization horizon approach that were identified with the copper plate model representation of the power systems discussed in the previous section. However, the 3-day extended window of foresight (ELH-3d) showed 1.8% and 1.3% improvement in terms of production cost when compared with the traditional dispatch method for the RTS-GMLC solar PV-driven and wind-driven systems, respectively. Similarly, the MT-based end volume target method (EVT-LA-MT) showed a 1.0% and 0.8% improvement in terms of production cost when compared with the traditional dispatch method for the RTS-GMLC solar PV-driven and wind-driven systems, respectively. Note that this performance is significantly lower than the performance of the ELH-3d dispatch method.

These improvements are significantly lower than the improvements observed with the copper plate model representation of the power systems, which could be partially driven by the fact that the transmission cases are not based on an optimal storage and VRE sizing and siting. Including that information, could increase the role and value of storage technologies and LDES dispatch approaches. Indeed, the goal of the transmission-constrained cases is not to assess the value of modeling LDES but the performance of the LDES dispatch approaches, including the scalability aspect. On the other hand, the flat SOC end volume target (EVT-LA) and the flat energy value (EV-01 and EV-025) target methods did not show promising or consistent performance in terms of production cost improvement, which is consistent with the results based on the copper plate power system representation. Regarding the standard capacity credit, both ELH-3d and EVT-LA-MT showed a superior performance compared with any other dispatch approach, including the traditional and the LP relaxation of the ideal case. Note that regardless of the energy mix, the SOC-aware peak net load-based capacity credit for LDES was always 100% for the extended optimization horizon and the end volume targets dispatch approaches. Moreover, for the ELH-3d dispatch approach, LDES provide between 43.7% and 75.3% of the total required reserves during the top 10 net load hours, depending on the energy mix. The LDES reserve provision is between 50.3% and 54.3% for the EVT-LA-MT dispatch approach. Thus, these two LDES dispatch methods are more effective at dispatching LDES for critical peak net load periods.

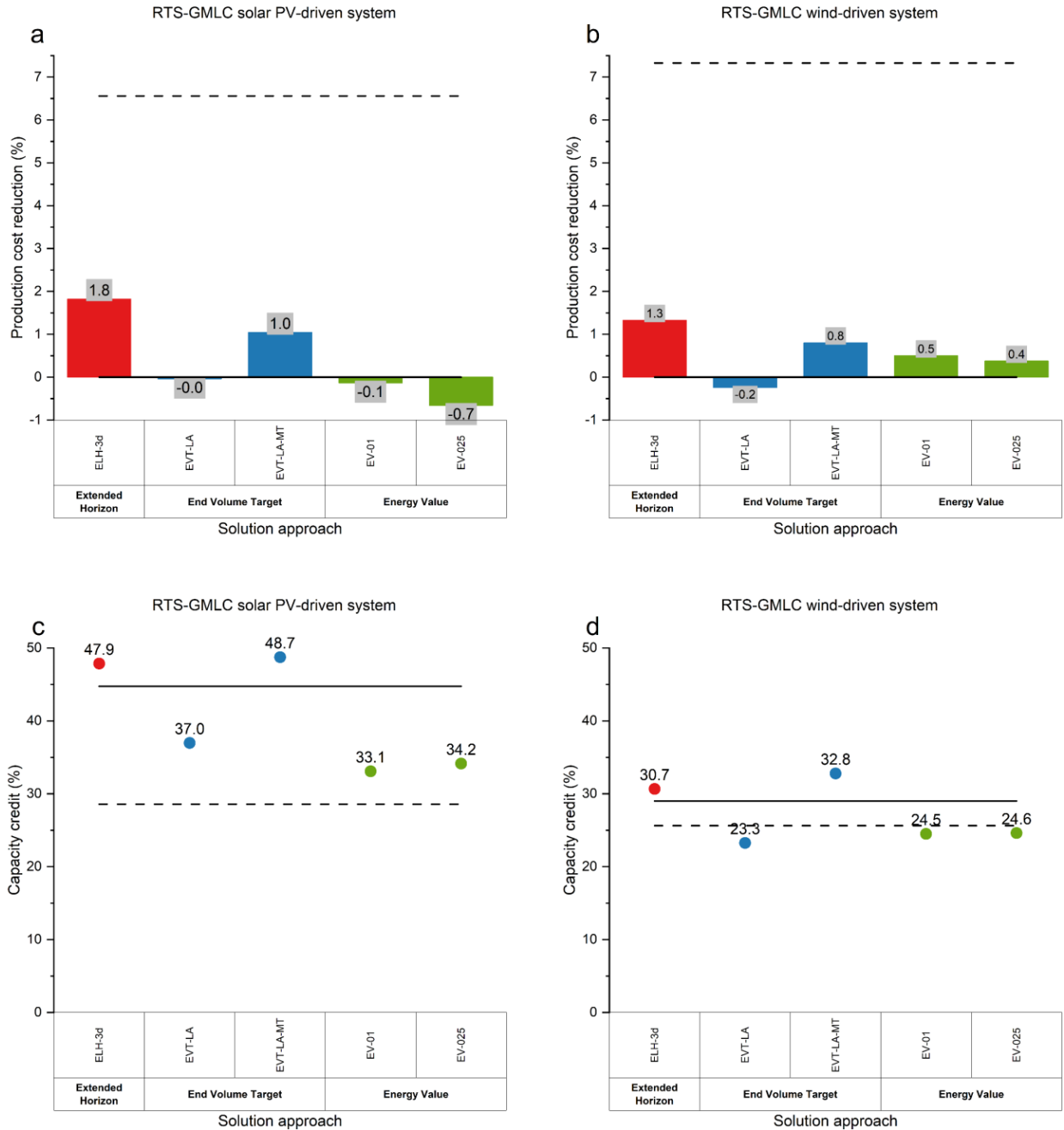

Figure 14. Production cost reduction (based on the traditional dispatch) (e.g., a and b), and standard capacity credit (e.g., c and d), for Reliability Test System Grid Modernization Lab Consortium (RTS-GMLC) solar PV- and wind-driven test power systems. The standard capacity credit was based on top 10 peak net load hours.

Regarding the scalability of the LDES dispatch approaches for transmission-constrained power systems, Figure 15 summarizes the CPU time and the peak memory usage for the LDES dispatch approaches. The CPU time for the 3-day extended window of foresight (ELH-3d) increases by more than 4-fold when compared with the traditional dispatch approach (for both solar PV- and wind-driven systems). This clearly illustrates the scalability issues associated with the extended optimization horizon LDES dispatch approach. Note that for the copper plate power system representation, the ELH-3d LDES dispatch approaches increase CPU time by less than 2-fold when compared with the traditional dispatch approach, regardless of the power system. Thus, the scalability of the LDES dispatch approach is more relevant for transmission-constrained power systems (nodal representation of the power systems). On the other hand, the end volume target and energy value dispatch approaches showed good performance in terms of scalability, e.g., CPU time and peak memory usage on the same scale as the traditional dispatch approach, which is consistent with the results based on the copper plate power system representation.

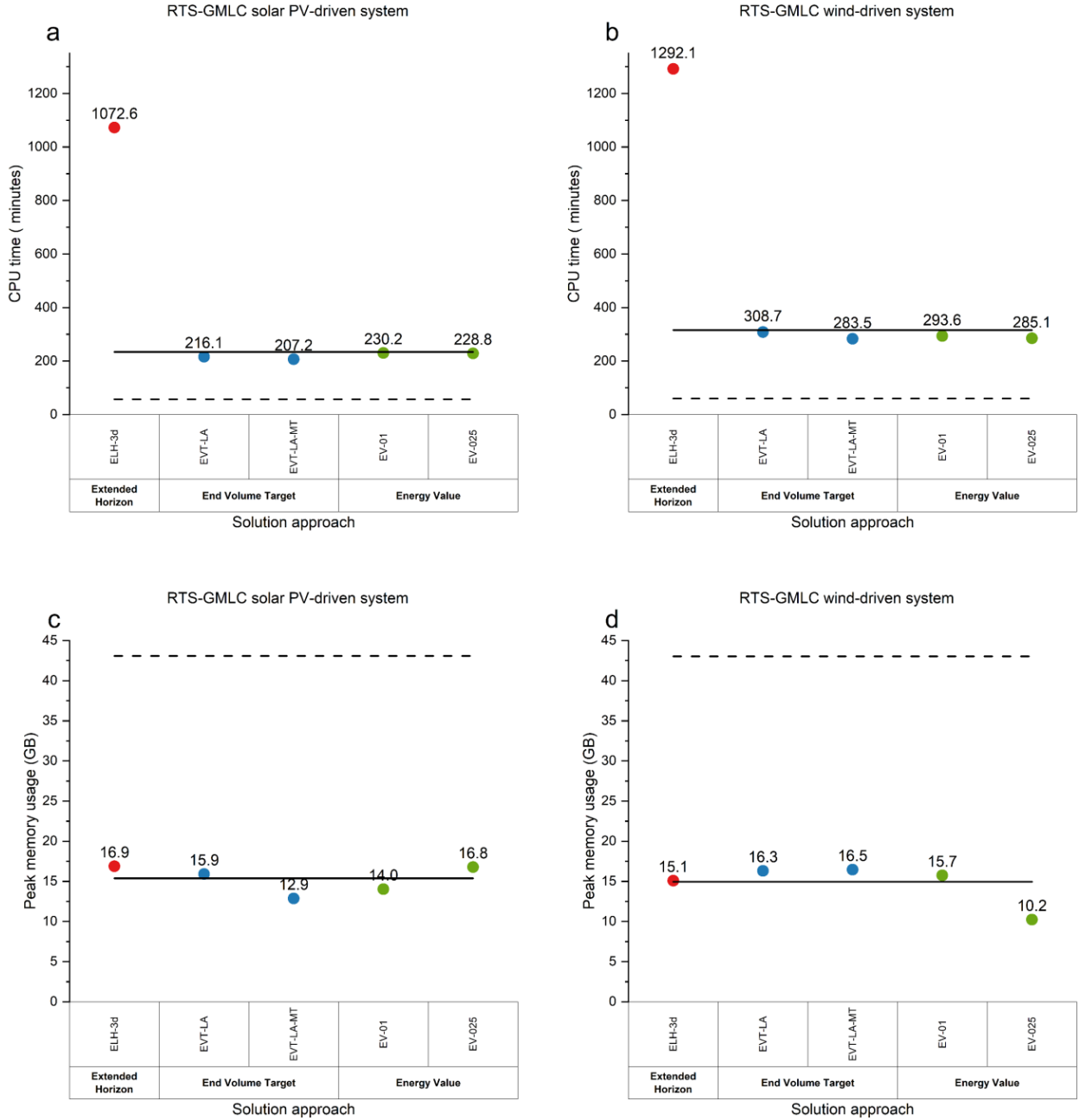

Figure 15. Central processing unit (CPU) time, e.g., (a) and (b), and peak memory usage, e.g., (b) and (c), for Reliability Test System Grid Modernization Lab Consortium (RTS-GMLC) solar PV- and wind-driven test power systems.

## 4.4 Effects of VRE uncertainty on the performance of long-duration energy storage dispatch approaches

Wind and solar PV resources are not only variable (e.g., power output varies across different timescales), but also uncertain given the stochastic nature of weather and climate variables. Thus, the uncertainty associated with VRE power output could have a significant impact on the operation of dispatchable generation/storage assets, including LDES. System operators usually perform VRE forecasts (e.g., day-ahead forecasting) to attempt to mitigate the impacts of VRE uncertainty in the operation of power grids with significant shares of wind and solar PV power sources. We tested the performance of the two LDES dispatch approaches (e.g., extended optimization horizon and end volume targets) in view of VRE generation uncertainty for the RTS-GMLC solar PV- and wind-driven test power systems. To this end, we created a synthetic day-ahead forecast for solar PV and wind generation by drawing from a normal error distribution with mean of error 3% and 6% respectively for each time period in the look-ahead window. For the RTS-GMLC solar PV-driven system, the system level normalized root mean square errors for the day-ahead forecasts are 13.0% for wind and 3.5% for solar PV, and the system level normalized mean absolute errors are 10.5% for wind and 2.2% for solar PV. Similarly, for the RTS-GMLC wind-driven system, the system level normalized root mean square errors for the day-ahead forecasts are 11.9% for wind and 3.4% for solar PV, and the system level normalized mean absolute errors are 9.5% for wind and 2.1% for solar PV. These day-ahead forecast errors are comparable with those from previous work[78,79].

Both copper plate and nodal representations of the RTS-GMLC test power system were tested in view of VRE generation uncertainty with the following LDES dispatch methods: extended window of foresight with 3 days look-ahead (ELH-3d) and the MT-based end volume target method (EVT-LA-MT). The day-ahead VRE forecast is used in different ways, depending on the dispatch method. For example, for the traditional 1 day-ahead plus 1 day look-ahead and ELH-3d dispatch approaches, the forecast is used for the look-ahead windows only (e.g., assuming perfect foresight for the 1 day-ahead window). For the EVT-LA-MT dispatch approach, the day-ahead forecast is used for the MT model that provides the end volume targets, while the traditional dispatch model assumes perfect foresight for the 1 day-ahead period and the VRE forecast for the 1-day look-ahead windows. With this accounting of uncertainty and for the copper plate power system representation, the LDES dispatch approaches could reduce production cost between 8.4% and 12.2% with respect to the traditional dispatch (see Table 2), which is comparable with the results for the copper plate representation with perfect foresight.

For the nodal representation (transmission constrained), the LDES dispatch approaches could reduce the production cost between 0.3% and 1.7%, with respect to the traditional dispatch approach (see Table 2). This is comparable to the results from the transmission constrained cases with perfect foresight (e.g., production cost reductions between 0.8% and 1.8% for the ELH-3d and EVT-LA-MT dispatch approaches). In general, the two LDES dispatch approaches increase the standard capacity credit of total energy storage capacity (combined short-duration and LDES) when compared with the traditional dispatch approach. Note that regardless of the energy mix, the SOC-aware peak net load-based capacity credit for LDES was greater than or equal to 84% for the ELH-3d and EVT-LA-MT dispatch approaches (see Table 2). Additionally, for the ELH-3d dispatch approach, LDES provide between 74.3% and 80.9% of the total required reserves during the top 10 net load hours, depending on the model formulation (copper plate or nodal representation) and the energy mix. In contrast, the LDES reserve provision is between 34.5% and 89.6% for the EVT-LA-MT dispatch approach. Depending on the model formulation (copper plate or nodal representation) and the energy mix, for the ELH-3d dispatch approach short-duration energy storage provides between 13.7% and 23.3% of the total required reserves during the top 10 net load hours. The short-duration energy storage reserve provision is between 3% and 54.3% for the EVT-LA-MT dispatch approach. These results, which assume perfect foresight for the day-ahead period and a VRE forecast for the look-ahead window, clearly demonstrate the value of more appropriate LDES dispatch approaches in view of VRE generation uncertainties. Also, considering VRE forecasts for the day-ahead period would likely further increase the value of LDES dispatch approaches. To fully capture the value of LDES dispatch approaches considering power system operational uncertainty, load, wind, and solar PV forecasts across multiple timescales should be used to provide the best information available to the dispatch algorithms.

Table 2. Production cost reduction and SOC-aware capacity credit for LDES dispatch approaches considering VRE uncertainty for the look-ahead and the RTS-GMLC tests power system.

| Case | System representation | Method | Production cost reduction (%) | LDES SOC-aware capacity credit (%) |
|---|---|---|---|---|
| Solar PV-driven | Copper plate | Traditional | - | 100 |
| | | ELH-3d | 10.8 | 100 |
| | | EVT-LA-MT | 12.2 | 100 |
| | Nodal | Traditional | - | 100 |
| | | ELH-3d | 1.7 | 100 |
| | | EVT-LA-MT | 0.7 | 100 |
| Wind-driven | Copper plate | Traditional | - | 87.3 |
| | | ELH-3d | 9.0 | 84.3 |
| | | EVT-LA-MT | 8.4 | 97.9 |
| | Nodal | Traditional | - | 100 |
| | | ELH-3d | 1.0 | 100 |
| | | EVT-LA-MT | 0.3 | 100 |

## 4.5 Future research directions

The LDES dispatch case studies and results presented in this study illustrate that both improved system representation and scalability should be considered when assessing LDES dispatch approaches. Indeed, only the MT-based end volume target method (EVT-LA-MT) showed promising performance in terms of improved system representation and scalability for both copper plate and nodal representations (transmission-constrained) of the power systems, as illustrated in Figure 16. Moreover, the estimated system value and computational requirements of the extended window of foresight increases with the length of the look-ahead horizon. Thus, scalability is an issue for this LDES dispatch method particularly for nodal representations of the power system (Figure 16). For example, moving from 1 day look-ahead to 3 days look-ahead (adding two additional days of foresight) increases the CPU time by more than 4-fold for the RTS-GMLC nodal model formulation. On the other hand, the energy value approach showed good performance in terms of scalability but mixed results in terms of improved system value of LDES.

Different research gaps remain regarding the dispatch modeling of LDES. First, the performance of the MT-based end volume target method depends strongly on how close the MT model represents the operation of the power system. Thus, this method could be improved by a better modeling of the flexibility provided by the thermal fleet (e.g., ramp, start up, and shutdown constraints). Additionally, using the end volume targets as a minimum SOC constraint could provide additional flexibility to accommodate future net load variations, while ensuring a minimum energy reserve that could be critical from a reliability standpoint. On the other hand, because scalability is the major limitation of the extended window of foresight approach, future research should focus on strategies to reduce the computation requirements for extending the look-ahead horizon. This could be achieved by performing time aggregation for the look-ahead horizon (e.g., 2 hours, 3 hours, or 4 hours aggregation for the look-ahead). Note that there is a trade-off between the length of the look-ahead horizon and the time aggregation (e.g., a longer look-ahead horizon will improve the system representation [estimated system value] but the time aggregation will reduce both the computational requirement as well as the estimated system value). Also, the required length of the look-ahead could depend on the duration of the LDES.

Thus, for a given power system, different numerical experiments should be performed in terms of the duration of the storage devices, the length of the look-ahead horizon and the time aggregation to identify the optimal settings for the extended window of foresight approach. Then, after identifying an adequate LDES dispatch approach, efforts should be devoted to enhancing the modeling of LDES in capacity and transmission expansion power system models. This will allow for better sizing and sitting of LDES in view of the integration of large shares of wind and solar PV power sources. Moreover, the deployment and dispatch of LDES in view of extreme climate or weather events or reliability applications has been largely overlooked. Therefore, more research efforts should be devoted towards to developing better modeling of LDES in view of extreme weather events or reliability applications, considering the associated uncertainties at both power system planning and operational levels. Indeed, the design and assessments of LDES dispatch methods should consider operational uncertainties associated with the variability of wind and solar PV power sources as well as electricity load.

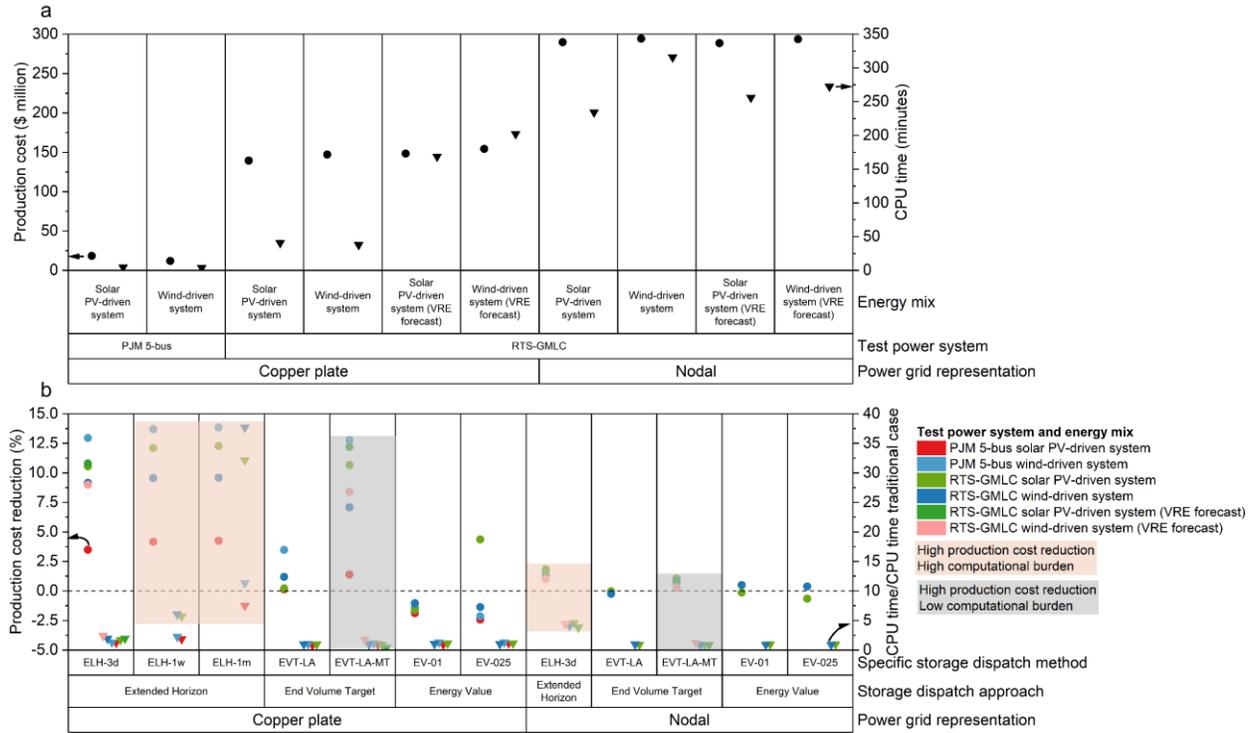

Figure 16. Production cost and CPU time for the traditional dispatch approach (a) and production cost reduction, e.g., with respect to the traditional dispatch approach, and normalized CPU time for the LDES dispatch approaches (b). Solid dots represent either production cost for traditional cases, e.g., black dots, (a) or production cost reduction for LDES dispatch approaches, e.g., colored dots based on test power system and energy mix, (b). Solid triangles represent either CPU time for traditional cases, e.g., black triangles, (a) or normalized CPU time for LDES dispatch approaches, e.g., colored triangles based on test power system and energy mix, (b). Production cost and production cost reduction are plotted in the left vertical axis of figures (a) and (b), respectively. CPU time and normalized CPU time are plotted in the right vertical axis of figures (a) and (b), respectively.

# 5 Conclusions

LDES devices could facilitate the integration of larger shares of VRE in power systems. Thus, the dispatch modeling of LDES could be critical for the development of integrated resource plans and the operation of high renewable power systems. This manuscript identifies the current state of the art for long-duration storage dispatch challenges and potential solutions. Although long-duration storage is expected to become more valuable to power system operations as shares of variable renewable generation increase, there has yet to be identified a broadly applicable method to accurately value these types of assets in traditional production cost modeling simulations. Such a method should be scalable, reasonably accurate, and relatively easy to implement within a standard power system production cost modeling framework. Existing methods may perform well for specific applications or for smaller systems, but these methods frequently suffer from computational challenges when attempting to scale up in spatial extent or to multiple storage devices.

From the literature we identified three relatively easy to implement LDES dispatch methods, including (i) extended optimization horizon, (ii) end volume targets, and (iii) stored energy value. Then, based on two test power systems considering both solar PV- and wind-driven renewable energy mixtures with 70%-90% VRE shares, we estimated the potential system value of modeling

LDES by comparing the optimal power system operation provided by the traditional dispatch approach (24 hours optimization window + 24 hours look-ahead horizon) with the ideal dispatch approach, (8,760 hours optimization window with no look-ahead). We estimated (e.g., based on the copper plate power system representation) that a better long-duration dispatch approach could increase the long-duration operational value (reduce the production cost of the power system) by 4% - 14%, depending on the test power system and the renewable energy mix. Additionally, we estimated that a better dispatch approach could increase the standard capacity credit by 14% - 34%, and therefore the capacity value, of LDES. Given the scale of power systems in the United States, the improvements in LDES operational and capacity values could represent significant production cost saving opportunities for system operators and electric utilities.

We then evaluated the performance of the three selected long-duration dispatch approaches in terms of both improved system value (e.g., based on production cost and standard capacity credit) and scalability (e.g., based on CPU time and peak memory usage) for both copper plate and nodal representations of the power system. The stored energy value approach did not show consistent performance across the cases evaluated in this study. Additionally, the extended optimization horizon showed good performance in terms of improved system value, but a poor performance in terms of scalability, e.g., the computational requirements increased exponentially with the length of the look-ahead horizon. The mid-term scheduling-based end volume target dispatch approach showed promising performance in terms of both improved system value and scalability. Thus, this method is attractive from a practical viewpoint. Based on these findings, we have identified a need for further analysis into optimal operation of long-duration storage. This analysis includes: (i) improving the performance of existing LDES dispatch methods, (ii) enhancing the modeling of LDES in capacity and transmission expansion power system models, and (iii) development of LDES dispatch approaches in view of extreme weather events or reliability applications, considering the associated uncertainties at both power system planning and operational levels.

# Acknowledgements

This work was authored in part by the National Renewable Energy Laboratory, operated by Alliance for Sustainable Energy, LLC, for the U.S. Department of Energy (DOE) under Contract No. DE-AC36-08GO28308. The views expressed in the article do not necessarily represent the views of the DOE or the U.S. Government. The U.S. Government retains and the publisher, by accepting the article for publication, acknowledges that the U.S. Government retains a nonexclusive, paid-up, irrevocable, worldwide license to publish or reproduce the published form of this work, or allow others to do so, for U.S. Government purposes.

# 6 References

1. U.S. Energy Information Agency (EIA). Renewable energy explained.

    https://www.eia.gov/energyexplained/renewable-sources/ (2021).

2. IEA. *Global Energy Review 2021*. https://www.iea.org/reports/global-energy-review-2021 (2021).

3. Denholm, P., Cole, W., Frazier, A. W., Podkaminer, K. & Blair, N. *The Four Phases of Storage Deployment: A Framework for the Expanding Role of Storage in the U.S. Power System*. https://www.nrel.gov/docs/fy21osti/77480.pdf (2021).

4. Augustine, C. & Blair, N. *Storage Futures Study: Storage Technology Modeling Input Data Report*. https://www.nrel.gov/docs/fy21osti/78694.pdf (2021).

5. Cole, W., Corcoran, S., Gates, N., Mai, T. & Das, P. *2020 Standard Scenarios Report: A U.S. Electricity Sector Outlook*. 51 (2020).

6. Vimmerstedt, L. *et al.* Annual Technology Baseline: The 2022 Electricity Update. (2022).

7. Blanco, H. & Faaij, A. A review at the role of storage in energy systems with a focus on Power to Gas and long-term storage. *Renewable and Sustainable Energy Reviews* **81**, 1049–1086 (2018).

8. Cole, W. J. *et al.* Quantifying the challenge of reaching a 100% renewable energy power system for the United States. *Joule* **5**, 1732–1748 (2021).

9. Hargreaves, J. J. & Jones, R. A. Long Term Energy Storage in Highly Renewable Systems. *Front. Energy Res.* **8**, 219 (2020).

10. Jafari, M., Korpås, M. & Botterud, A. Power system decarbonization: Impacts of energy storage duration and interannual renewables variability. *Renewable Energy* **156**, 1171–1185 (2020).

11. de Sisternes, F. J., Jenkins, J. D. & Botterud, A. The value of energy storage in decarbonizing the electricity sector. *Applied Energy* **175**, 368–379 (2016).

12. Sepulveda, N. A., Jenkins, J. D., Edington, A., Mallapragada, D. S. & Lester, R. K. The design space for long-duration energy storage in decarbonized power systems. *Nat Energy* **6**, 506–516 (2021).

13. Sioshansi, R., Madaeni, S. H. & Denholm, P. A Dynamic Programming Approach to Estimate the Capacity Value of Energy Storage. *IEEE Transactions on Power Systems* **29**, 395–403 (2014).

14. Dowling, J. A. *et al.* Role of Long-Duration Energy Storage in Variable Renewable Electricity Systems. *Joule* **4**, 1907–1928 (2020).

15. Denholm, P., Cole, W., Frazier, A. W., Podkaminer, K. & Blair, N. *The Challenge of Defining Long-Duration Energy Storage*. https://www.nrel.gov/docs/fy22osti/80583.pdf (2021).

16. Tuttman, M. & Litzelman, S. Why Long-Duration Energy Storage Matters. *Advanced Research Projects Agency-Energy (ARPA-E)* https://arpa-e.energy.gov/news-and-media/blog-posts/why-long-duration-energy-storage-matters (2020).

17. Gaffney, F., Deane, J. P., Drayton, G., Glynn, J. & Gallachóir, B. P. Ó. Comparing negative emissions and high renewable scenarios for the European power system. *BMC Energy* **2**, 3 (2020).

18. Pellow, M., Eichman, J., Zhang, J. & Guerra, O. J. *Valuation of Hydrogen Technology on the Electric Grid Using Production Cost Modeling*. (2021).

19. de Guibert, P., Shirizadeh, B. & Quirion, P. Variable time-step: A method for improving computational tractability for energy system models with long-term storage. *Energy* **213**, 119024 (2020).

20. Sioshansi, R. *et al.* Energy-Storage Modeling: State-of-the-Art and Future Research Directions. *IEEE Trans. Power Syst.* 1–1 (2021) doi:10.1109/TPWRS.2021.3104768.

21. Guerra, O. J. *et al.* The value of seasonal energy storage technologies for the integration of wind and solar power. *Energy & Environmental Science* **13**, 1909–1922 (2020).

22. Götz, M. *et al.* Renewable Power-to-Gas: A technological and economic review. *Renewable Energy* **85**, 1371–1390 (2016).

23. Sioshansi, R. & Denholm, P. The value of concentrating solar power and thermal energy storage. *IEEE Transactions on Sustainable Energy* **1**, 173–183 (2010).

24. Guerra, O. J., Eichman, J. & Denholm, P. Optimal energy storage portfolio for high and ultrahigh carbon-free and renewable power systems. *Energy & Environmental Science* **14**, 5132–5146 (2021).

25. Jorgenson, J., Frazier, A. W., Denholm, P. & Blair, N. *Storage Futures Study: Grid Operational Impacts of Widespread Storage Deployment*. https://www.nrel.gov/docs/fy22osti/80688.pdf (2022).

26. Guerra, O. J. Beyond short-duration energy storage. *Nat Energy* **6**, 460–461 (2021).

27. Bistline, J. *et al.* Energy storage in long-term system models: a review of considerations, best practices, and research needs. *Progress in Energy* **2**, 032001 (2020).

28. Cole, W., Denholm, P., Carag, V. & Frazier, W. The peaking potential of long-duration energy storage in the United States power system. *Journal of Energy Storage* **62**, 106932 (2023).

29. Zhang, J., Guerra, O. J., Eichman, J. & Pellow, M. A. Benefit Analysis of Long-Duration Energy Storage in Power Systems with High Renewable Energy Shares. *Front. Energy Res.* **8**, 527910 (2020).

30. Sioshansi, R., Denholm, P., Jenkin, T. & Weiss, J. Estimating the value of electricity storage in PJM: Arbitrage and some welfare effects. *Energy Economics* **31**, 269–277 (2009).

31. Jorgenson, J., Denholm, P. & Mehos, M. *Estimating the Value of Utility-Scale Solar Technologies in California Under a 40% Renewable Portfolio Standard*. https://www.nrel.gov/docs/fy14osti/61685.pdf (2014).

32. Jorgenson, J., Denholm, P., Mehos, M. & Turchi, C. *Estimating the Performance and Economic Value of Multiple Concentrating Solar Power Technologies in a Production Cost Model*. http://www.nrel.gov/docs/fy14osti/58645.pdf (2013).

33. Zhang, Z., Zhang, Y. & Lee, W.-J. Energy storage based optimal dispatch scheme for financial improvement and fluctuation mitigation on wind power generation. in *2017 IEEE Industry Applications Society Annual Meeting* 1–7 (2017). doi:10.1109/IAS.2017.8101728.

34. Schill, W.-P. & Kemfert, C. Modeling Strategic Electricity Storage: The Case of Pumped Hydro Storage in Germany. *EJ* **32**, (2011).

35. Petrollese, M., Seche, P. & Cocco, D. Analysis and optimization of solar-pumped hydro storage systems integrated in water supply networks. *Energy* **189**, 116176 (2019).

36. Denholm, P. *et al.* The value of energy storage for grid applications. *Contract* **303**, 275–3000 (2013).

37. Cebulla, F., Naegler, T. & Pohl, M. Electrical energy storage in highly renewable European energy systems: Capacity requirements, spatial distribution, and storage dispatch. *Journal of Energy Storage* **14**, 211–223 (2017).

38. DOE *et al. Solar Futures Study*. 310 (2021).

39. Cebulla, F., Haas, J., Eichman, J., Nowak, W. & Mancarella, P. How much electrical energy storage do we need? A synthesis for the U.S., Europe, and Germany. *Journal of Cleaner Production* **181**, 449–459 (2018).

40. Balducci, P. J., Alam, M. J. E., Hardy, T. D. & Wu, D. Assigning value to energy storage systems at multiple points in an electrical grid. *Energy Environ. Sci.* **11**, 1926–1944 (2018).

41. Haas, J. *et al.* Challenges and trends of energy storage expansion planning for flexibility provision in low-carbon power systems – a review. *Renewable and Sustainable Energy Reviews* **80**, 603–619 (2017).

42. Lai, C. S., Locatelli, G., Pimm, A., Wu, X. & Lai, L. L. A review on long-term electrical power system modeling with energy storage. *Journal of Cleaner Production* **280**, 124298 (2021).

43. Sánchez-Pérez, P. A., Staadecker, M., Szinai, J., Kurtz, S. & Hidalgo-Gonzalez, P. Effect of modeled time horizon on quantifying the need for long-duration storage. *Applied Energy* **317**, 119022 (2022).

44. Tejada-Arango, D. A., Domeshek, M., Wogrin, S. & Centeno, E. Enhanced Representative Days and System States Modeling for Energy Storage Investment Analysis. *IEEE Transactions on Power Systems* **33**, 6534–6544 (2018).

45. Zurita, A., Mata-Torres, C., Cardemil, J. M., Guédez, R. & Escobar, R. A. Multi-objective optimal design of solar power plants with storage systems according to dispatch strategy. *Energy* **237**, 121627 (2021).

46. Koko, S. P., Kusakana, K. & Vermaak, H. J. Optimal power dispatch of a grid-interactive micro-hydrokinetic-pumped hydro storage system. *Journal of Energy Storage* **17**, 63–72 (2018).

47. Deane, J. P., McKeogh, E. J. & Gallachoir, B. P. O. Derivation of Intertemporal Targets for Large Pumped Hydro Energy Storage With Stochastic Optimization. *IEEE Transactions on Power Systems* **28**, 2147–2155 (2013).

48. Hittinger, E. & Ciez, R. E. Modeling Costs and Benefits of Energy Storage Systems. *Annual Review of Environment and Resources* **45**, 445–469 (2020).

49. Brinkman, G., Novacheck, J. & Ho, J. *The North American Renewable Integration Study (NARIS): A U.S. Perspective*. (2021).

50. Jorgenson, J., Hale, E. & Cowiestoll, B. *Managing Solar Photovoltaic Integration in the Western United States: Power System Flexibility Requirements and Supply*. (2020).

51. Steinberg, D. *et al. Chapter 6. Renewable Energy Investments and Operations*. 216 https://www.nrel.gov/docs/fy21osti/79444-6.pdf (2021).

52. Niet, T. Storage end effects: An evaluation of common storage modelling assumptions. *Journal of Energy Storage* **27**, 101050 (2020).

53. Henze, G. P., Dodier, R. H. & Krarti, M. Development of a Predictive Optimal Controller for Thermal Energy Storage Systems. *HVAC&R Research* **3**, 233–264 (1997).

54. Helseth, A. *et al.* Hydropower Scheduling Toolchains: Comparing Experiences in Brazil, Norway, and USA and Implications for Synergistic Research. *Journal of Water Resources Planning and Management* **149**, (2023).

55. Efthymoglou, P. G. Optimal Use and the Value of Water Resources in Electricity Generation. *Management Science* **33**, 1622–1634 (1987).

56. Ferreira, L. a. F. M. Short-term scheduling of a pumped storage plant. *IEE Proceedings C (Generation, Transmission and Distribution)* **139**, 521–528 (1992).

57. Dasigenis, A. T. & Garcia-San Pedro, A. R. Real-time hydro coordination and economic hydro optimization. in *Proceedings of Power Industry Computer Applications Conference* 150–157 (1995). doi:10.1109/PICA.1995.515178.

58. Foley, A. & Díaz Lobera, I. Impacts of compressed air energy storage plant on an electricity market with a large renewable energy portfolio. *Energy* **57**, 85–94 (2013).

59. Cochran, J. & Denholm, P. *The Los Angeles 100% Renewable Energy Study*. (2021).

60. Teichgraeber, H. & Brandt, A. R. Time-series aggregation for the optimization of energy systems: Goals, challenges, approaches, and opportunities. *Renewable and Sustainable Energy Reviews* **157**, 111984 (2022).

61. Orwig, K. D. *et al.* Recent Trends in Variable Generation Forecasting and Its Value to the Power System. *IEEE Transactions on Sustainable Energy* **6**, 924–933 (2015).

62. Sergi, B. *et al. ARPA-E PERFORM datasets*. (2022).

63. Padhy, N. P. Unit Commitment—A Bibliographical Survey. *IEEE Trans. Power Syst.* **19**, 1196–1205 (2004).

64. Abujarad, S. Y., Mustafa, M. W. & Jamian, J. J. Recent approaches of unit commitment in the presence of intermittent renewable energy resources: A review. *Renewable and Sustainable Energy Reviews* **70**, 215–223 (2017).

65. IRENA. *Planning for the Renewable Future: Long-term modelling and tools to expand variable renewable power in emerging economies*. (International Renewable Energy Agency (IRENA), 2017).

66. Davies, D. M. *et al.* Combined economic and technological evaluation of battery energy storage for grid applications. *Nat Energy* **4**, 42–50 (2019).

67. Pandzic, H. & Bobanac, V. An Accurate Charging Model of Battery Energy Storage. *IEEE Trans. Power Syst.* **34**, 1416–1426 (2019).

68. Gür, T. M. Review of electrical energy storage technologies, materials and systems: challenges and prospects for large-scale grid storage. *Energy Environ. Sci.* **11**, 2696–2767 (2018).

69. Beaudin, M., Zareipour, H., Schellenberglabe, A. & Rosehart, W. Energy storage for mitigating the variability of renewable electricity sources: An updated review. *Energy for Sustainable Development* **14**, 302–314 (2010).

70. Fangxing Li & Rui Bo. Small test systems for power system economic studies. in *IEEE PES General Meeting* 1–4 (IEEE, 2010). doi:10.1109/PES.2010.5589973.

71. Subcommittee, P. M. IEEE Reliability Test System. *IEEE Transactions on Power Apparatus and Systems* **PAS-98**, 2047–2054 (1979).

72. Barrows, C. *et al.* The IEEE Reliability Test System: A Proposed 2019 Update. *IEEE Transactions on Power Systems* **35**, 119–127 (2020).

73. Li, H., Sun, J. & Tesfatsion, L. Testing Institutional Arrangements via Agent-Based Modeling: A U.S. Electricity Market Application. in *Computational Methods in Economic Dynamics* (eds. Dawid, H. & Semmler, W.) vol. 13 135–158 (Springer Berlin Heidelberg, 2011).

74. Gonzalez-Fernandez, R. A., Leite da Silva, A. M., Resende, L. C. & Schilling, M. T. Composite Systems Reliability Evaluation Based on Monte Carlo Simulation and Cross-Entropy Methods. *IEEE Trans. Power Syst.* **28**, 4598–4606 (2013).

75. Bezanson, J., Edelman, A., Karpinski, S. & Shah, V. B. Julia: A Fresh Approach to Numerical Computing. *SIAM Rev.* **59**, 65–98 (2017).

76. Twitchell, J., DeSomber, K. & Bhatnagar, D. Defining long duration energy storage. *Journal of Energy Storage* **60**, 105787 (2023).

77. Jorgenson, J., Awara, S., Stephen, G. & Mai, T. *Comparing Capacity Credit Calculations for Wind: A Case Study in Texas*. https://www.nrel.gov/docs/fy21osti/80486.pdf (2021).

78. Hodge, B.-M. *et al.* The combined value of wind and solar power forecasting improvements and electricity storage. *Applied Energy* **214**, 1–15 (2018).

79. Guerra, O. J. *et al.* Coordinated operation of electricity and natural gas systems from day-ahead to real-time markets. *Journal of Cleaner Production* **281**, 124759 (2021).